\documentclass[lettersize,journal,twoside]{IEEEtran}
\IEEEoverridecommandlockouts

\usepackage{cite}
\usepackage{amsmath,amssymb,amsfonts}
\usepackage{algorithmic}
\usepackage{graphicx}
\usepackage{textcomp}
\usepackage{xcolor}
\usepackage[utf8]{inputenc} 
\usepackage[T1]{fontenc}
\usepackage{multirow}
\usepackage{booktabs} 
\usepackage[table]{xcolor} % Enables cell coloring
\usepackage{array}
\usepackage{graphicx}      % for \resizebox
\usepackage[caption=false,font=normalsize,labelfont=sf,textfont=sf]{subfig}
\usepackage{xurl} 
\usepackage{orcidlink}
\usepackage{stfloats}
\usepackage{tikz}
\usetikzlibrary{patterns}
\usepackage{eso-pic}

\newcommand{\IEEEFirstPageCopyright}{%
\AddToShipoutPictureFG*{%
  \AtPageLowerLeft{%
    \raisebox{0.20in}{%
      \makebox[\paperwidth][c]{%
        \begin{minipage}{0.94\paperwidth}
        \centering
        \footnotesize
        1556-6021~\copyright~2026 IEEE. All rights reserved, including rights for text and data mining, and training of artificial intelligence and\\
        similar technologies. Personal use is permitted, but republication/redistribution requires IEEE permission.\\
        See https://www.ieee.org/publications/rights/index.html for more information.
        \end{minipage}
      }%
    }%
  }%
}%
}

\tikzset{crosshatch/.style={pattern=north east lines, pattern color=black!50}}

\begin{document}

\newcommand{\znote}[1]{{\color{red}{\bf [#1]}}}
\newcommand{\rnote}[1]{{\color{blue}{\bf [#1]}}}
\newcommand{\change}[1]{{{#1}}}

\newcommand{\badcell}[1]{\cellcolor{red!25} \fcolorbox{red}{red!25}{#1}}

\title{AdvScan: Black-Box Adversarial Example Detection at Runtime through Power Analysis}

\author{Robi Paul\orcidlink{0000-0002-7367-8432}, \textit{Student Member, IEEE}, and Michael Zuzak\orcidlink{0000-0003-0356-9393}, \textit{Member, IEEE}

\thanks{Received 21 May 2025; revised 3 December 2025 and 29 January 2026; accepted 1 February 2026. Date of publication 9 February 2026; date of current version 18 February 2026. The associate editor coordinating the review of this article and approving it for publication was Dr. Stjepan Picek. \textit{(Corresponding author: Robi Paul.)}

The authors are with Rochester Institute of Technology, Rochester, NY 14623 USA (e-mail: rp7248@rit.edu; mjzeec@rit.edu).

Digital Object Identifier 10.1109/TIFS.2026.3663053}}

% The paper headers
\markboth{IEEE Transactions on Information Forensics and Security}%
{Paul And Zuzak: AdvScan: Black-Box AE Detection at Runtime through Power Analysis}

\maketitle
\IEEEFirstPageCopyright

\begin{abstract}
TinyML models deployed on edge devices are increasingly adopted in safety/security-critical applications, making them a prime target for adversarial example (AE) attacks where inputs are modified to cause misclassifications. However, existing AE detection methods either require white-box model access, which is often unavailable in licensed black-box deployments, or rely on input pre-processing stages that add non-trivial latency and resource overhead, often exceeding what mission-critical applications can afford on their inference path. To address these challenges, we propose AdvScan, a runtime power analysis-based methodology for AE detection that operates in a black-box scenario and while inducing minimal latency. AdvScan is based on the observation that AEs produce anomalous neuron activations, which, in turn, generate distinctive power‐consumption signatures. The algorithm initially constructs a baseline distribution of power signatures from known benign inputs, then, at runtime, applies a one-sample t-test to determine whether a test input’s power signature significantly deviates from this baseline, thereby detecting AEs. We evaluated AdvScan using three adversarial example (AE) generation algorithms (Fast Gradient Sign Method (FGSM), Projected Gradient Descent (PGD), and Carlini–Wagner (C\&W)) on three MLPerf Tiny benchmark models implemented on two target devices: the STM32F303RC (ARM Cortex-M4) and STM32L562RE (ARM Cortex-M33) microcontrollers. Across 318,400 total test inputs, AdvScan detects 99.984\% of AEs with only 40 false negatives and zero false positives. These results demonstrate the viability of power-based AE detection for secure, accuracy-critical TinyML deployments in black-box environments.
\end{abstract}


\section{Introduction}

\IEEEPARstart{N}{eural} networks (NNs) have become a cornerstone of edge computing by enabling devices to perform complex tasks with reduced latency while preserving user privacy. Rising demands for efficiency, autonomy, and security have accelerated the deployment of NNs on edge devices \cite{abadade2023comprehensive, im2024tinyml}, fostering the development of \textit{TinyML} models optimized for ultra-low power and resource-constrained environments \cite{banbury2020benchmarking}. However, the simplicity of such models, coupled with challenging deployment conditions, makes them particularly vulnerable to attacks \cite{goodfellow2014explaining, mkadry2017towards, carlini2017towards}.

Adversarial perturbations deliberately mislead NN models through subtle modifications to the input data (i.e., AEs) \cite{akhtar2021advances, goodfellow2014explaining, mkadry2017towards, carlini2017towards} or by altering the internal model parameters at the user end \cite{gu2019badnets, kravchik2021poisoning} to achieve specific adversarial goals with minimal accuracy loss for benign inputs. To generate AEs, attackers typically use optimization-based techniques, such as the Fast Gradient Sign Method (FGSM) \cite{goodfellow2014explaining}, Projected Gradient Descent (PGD) \cite{mkadry2017towards}, and Carlini-Wagner (C\&W) attacks \cite{carlini2017towards}, to maximize prediction errors while maintaining stealthy perturbations. The potential consequences of such attacks can be severe, especially in critical applications such as autonomous vehicles, where attacks may cause the vehicle's perception systems to fail in identifying pedestrians, road signs, or other vehicles \cite{cheng2023fusion}; healthcare systems, where adversarial inputs can lead to incorrect medical diagnoses and treatment decisions \cite{ma2021understanding}; and biometric security systems, allowing unauthorized parties to access sensitive data.

Considerable prior research has explored defense techniques, including adversarial re-training, which retrains models with known AEs \cite{jia2024improving, kuang2024defense}, and stochastic approaches that introduce inference-time randomness \cite{he2019parametric, wang2023addition, roth2019odds}. However, these defenses often fail against unknown or sophisticated AE attacks \cite{goodfellow2014explaining, mkadry2017towards, carlini2017towards}. Another research direction focuses on the external input pre-processing, employing input transformations such as feature denoising \cite{huang2024adversarial, zhu2024pixeldenoise}, and MagNet \cite{meng2017magnet}, reverting adversarial images with generative models \cite{zhang2021defense, ji2023purifying}. {Although these techniques can operate in black-box scenarios by pre-processing inputs prior to inference, they introduce additional computation, memory requirements, and latency overhead that can be difficult to accommodate in mission-critical edge deployments with tight constraints.} Moreover, in licensing scenarios, model owners may restrict end-users to black-box access to protect intellectual property (IP) \cite{leroux2022tinymlops}, prohibiting the application of white-box defense strategies entirely.

Despite advances in AE detection, a substantial gap remains for TinyML models in edge computing environments. Existing runtime detection approaches \cite{feinman2017detecting, pertigkiozoglou2018detecting} depend on white-box assumptions about model internals, making them unsuitable for black-box deployments (e.g., a licensed model). Additionally, input transformation methods \cite{huang2024adversarial, zhu2024pixeldenoise, meng2017magnet} typically rely on repeated pre-processing steps or auxiliary models, introducing non-trivial computational, memory, and latency requirements that are difficult to accommodate under strict edge constraints. Addressing this gap forms the central motivation of this work: \textbf{to develop a mathematically robust and efficient runtime AE detection algorithm for TinyML models, deployable in purely black-box, memory-constrained environments without compromising model accuracy}.

\subsection{Contributions}

In this work, we propose \textbf{AdvScan}, a runtime AE detection algorithm for TinyML models on resource-constrained edge devices. AdvScan operates in a black-box setting, requiring no trust or cooperation between the model owner and the end user. At its core, AdvScan builds on the observation that AEs trigger anomalous neuron activations in NNs \cite{liu2019abs}. Since a microcontroller unit's (MCU) power consumption typically correlates with computational activity, any adversarial behavior could deviate from expected benign characteristics, potentially resulting in distinguishable power signatures during inference. Leveraging this, AdvScan combines power analysis with statistical modeling and hypothesis testing to determine whether an input is adversarial. The primary contributions of this work are summarized below:

\begin{enumerate}
\item We develop \textbf{AdvScan}, a runtime AE detection method leveraging correlational power analysis to detect adversarial inputs in resource-constrained TinyML models deployed in black-box scenarios. Unlike existing methods, AdvScan operates without affecting model accuracy or requiring adversarial retraining.
\item We develop a mathematically robust detection algorithm for AdvScan based on hypothesis testing (one-sample t-test). This algorithm provides quantifiable detection confidence for an input to be an AE and can be tuned to meet user-preferred detection goals.
\item We evaluate AdvScan on {two MCU platforms, an STM32F303RC (Arm Cortex-M4) and an STM32L562RE (Arm Cortex-M33), running three TinyML models in the presence of AEs generated by Fast Gradient Sign Method (FGSM) \cite{goodfellow2014explaining}, Projected Gradient Descent (PGD) \cite{mkadry2017towards}, and the Carlini–Wagner (C\&W) attack \cite{carlini2017towards} under $L_0$, $L_2$, and $L_{\infty}$ norms. Across 318,400 test cases, comprising both benign inputs and AEs, AdvScan, with a user-defined detection threshold of $P_{th} = 10^{-5}$, at runtime correctly classifies 318,360 cases (99.987\% overall detection accuracy), with only 40 false negatives and zero false positives.} 
% \item We evaluate the performance of AdvScan on an STM32F303RC MCU (ARM Cortex-M4) running three TinyML models in the presence of AEs generated by three common algorithms: Fast Gradient Sign Method (FGSM) \cite{goodfellow2014explaining}, Projected Gradient Descent (PGD) \cite{mkadry2017towards}, and Carlini-Wagner (C\&W) Attack \cite{carlini2017towards}. Across 46,153 evaluated test cases, AdvScan detects 99.99\% of AEs with only four false negatives and zero false positives at runtime with a user-defined threshold of $(P_{th} < 10^{-5})$.
\end{enumerate}

\section{Preliminaries}

\subsection{Adversarial Examples in Neural Networks}
\label{sec:Adv_ex}
An adversarial example (AE) is a maliciously crafted input designed to make NN models misclassify. For example, consider a model, $y=F(x)$ that maps inputs $x \in X$ to output labels $y \in Y$. An attacker generates an AE $x'$ by adding a perturbation $\delta$ to a benign input $x$ such that $x' = x + \delta$. This perturbation ($\delta$) forces the model to produce an incorrect output (i.e., $F(x') \neq F(x)$), which can be either targeted (aiming for a specific incorrect label) or untargeted. In this work, we focus on three common AE generation algorithms: Fast Gradient Sign Method (FGSM) \cite{goodfellow2014explaining}, Projected Gradient Descent (PGD) \cite{mkadry2017towards}, and the Carlini–Wagner (C\&W) Attack \cite{carlini2017towards}. We briefly outline each method below.

\textbf{Fast Gradient Sign Method (FGSM)} generates AEs by computing a perturbation $\delta$ in the direction that maximizes the model's loss function $L$ and adding it to the original input  $x$ \cite{goodfellow2014explaining}. Formally, FGSM is defined in Eq. \ref{fgsm}, where $\epsilon$ is a user-selected parameter controlling the magnitude of the $\delta$. Here, the parameter $\eta \in \{-1, +1\}$ determines whether the attack is untargeted $(\eta =  +1)$ or targeted $(\eta =  -1)$. Although FGSM is computationally efficient due to its single-step nature, its simplicity also limits its overall effectiveness against more robust models \cite{hemashree2024enhancing}.

\begin{equation}
\label{fgsm}
    x' = x + \eta \cdot \epsilon\cdot \text{sign} \left( \nabla_x L(F(x), y') \right)
\end{equation}

\textbf{Projected Gradient Descent (PGD)} generates AEs by iteratively adding $\delta$ in the direction that maximizes the model’s loss function ($L$) while projecting the modified input back into a valid range (i.e., within $x \pm \epsilon$ under the $L_{\infty}$ norm) \cite{mkadry2017towards}. Formally, each iteration computes the gradient of $L$ for the current modified input and takes a step of size $\alpha$ in that direction, as shown in Eq. \ref{pgd}. The updated input is then clipped (projected, $\Pi$) so that $\|x_{t+1} - x\|_\infty \leq \epsilon$. Because PGD employs multiple iterative updates (in contrast to the single-step FGSM), it is generally more effective against robust models. However, this improved effectiveness comes at the cost of higher computational complexity. As with FGSM, $\epsilon$ is a user-controlled parameter determining the magnitude of $\delta$.

\begin{equation}
\label{pgd}
x_{t+1} = \Pi_{\|x_{t+1} - x\|_\infty \leq \epsilon} \left( x_t + \alpha \cdot \text{sign} \left( \nabla_{x_t} L(x_t, y) \right) \right)
\end{equation}

\textbf{Carlini-Wagner (C\&W)} formulates AE generation as an optimization problem that minimizes the distance (measured in an  $L_p$ norm) between the benign input  $x$ and its adversarial counterpart $x'$ while forcing misclassification \cite{carlini2017towards} (see Eq. \ref{cw}). Its objective function balances a distortion term  $\|x' - x\|_p$ with a classification loss $L(F(x'),y)$ via the constant $c$, and enforces $x' \in [0,1]^n$. Compared to gradient-based methods such as FGSM and PGD, the C\&W attack can be more effective against defenses, though it requires significantly more computation. In this study, we considered the objective function defined in Eq. \ref{cw_ob}.

\begin{equation}
\label{cw}
\min_{x'} \; \|x' - x\|_p + c \cdot L(F(x'), y)
\quad \text{subject to,} \  x' \in [0,1]^n
\end{equation}

\begin{equation}
    \label{cw_ob}
    f_1(x') = \max\left(\max_{i \neq l} F_i(x') - F_l(x'), 0\right)
\end{equation}

\subsection{AE Detection in Neural Networks}
Detecting AEs in TinyML models can be broadly classified into two categories based on the degree of model access. 

\textbf{White-box} detection requires direct access to the model’s internal architecture and parameters to detect input AEs. They detect possible adversarial modifications by analyzing intermediate features, hidden-layer activations, or the statistical distributions of model layers. For instance, uncertainty-based methods use Monte Carlo dropout to estimate Bayesian uncertainty \cite{feinman2017detecting}, while softmax or logits-based methods track the model's output confidence levels to identify anomalous behavior \cite{pertigkiozoglou2018detecting}. Gradient-based detectors (e.g., GraN \cite{lust2020gran}) inspect gradient norms for the input, revealing suspicious neuron activations. Statistical techniques such as Principal Component Analysis (PCA) \cite{li2017adversarial} and Kernel Density Estimation (KD) \cite{zhao2023adversarial} pinpoint minor shifts in internal activations triggered by AEs. While these methods are highly effective in detecting adversarial behavior in inputs, they become impractical in a licensing scenario, where the models are in black-box settings.

% \textbf{Black-box} detection methodologies assume no access to the model’s architecture, gradients, or parameters. Instead, they rely on observable input-output behaviors or auxiliary signals to detect adversarial inputs. Some approaches focus on input pre-processing (e.g., feature denoising \cite{huang2024adversarial, zhu2024pixeldenoise, wang2023addition}, MagNet \cite{meng2017magnet}), which removes or corrects perturbations before inference. While these techniques do not require trust or cooperation from the model owner, they risk degrading model accuracy or introducing computational overhead. However, many of these methods degrade benign accuracy or require extensive retraining, which limits their suitability for constrained edge deployments.

\textbf{Black-box} detection methodologies assume no access to the model’s architecture, gradients, or parameters. Instead, they rely on observable input-output behaviors or auxiliary signals to detect adversarial inputs. Some approaches focus on input pre-processing (e.g., feature denoising \cite{huang2024adversarial, zhu2024pixeldenoise, wang2023addition}, MagNet \cite{meng2017magnet}), which removes or corrects perturbations before inference. While these techniques do not require trust or cooperation from the model owner, {they often inadvertently discard important features from benign images. Compression-based detection methods \cite{xu2017feature} are another notable class of black-box approaches; however, they typically rely on additional forward passes over transformed inputs, thereby introducing computational overhead and detection delay.}

Although both white-box and black-box adversarial detection methods can be effective, {they are not well suited to} black-box, license-based TinyML in edge deployments, where model internals remain proprietary, and owner cooperation is not guaranteed. White-box techniques are impractical due to inaccessible internals, while black-box methods, such as input pre-processing, may degrade performance \cite{li2021cleanml}. To overcome these challenges, we propose a new approach that (1) operates in a black-box setting, (2) operates under stringent resources, and (3) requires no owner cooperation, thereby ensuring robust adversarial defense in resource-constrained TinyML environments.
% Other black-box methods, such as EMShepherd \cite{ding2023emshepherd}, leverage electromagnetic (EM) side-channel leakage and auxiliary classifiers trained to distinguish between benign and adversarial patterns in the observed EM traces and can detect AEs without requiring internal model details.

% Although both white-box and black-box adversarial detection methods can be effective, neither is fully suited to accuracy-critical license-based TinyML deployments, where model internals remain proprietary and owner cooperation is not guaranteed. White-box techniques are impractical due to inaccessible internals, while black-box methods, such as input pre-processing, may degrade performance \cite{li2021cleanml}. To overcome these challenges, we propose a new approach that (1) operates in a black-box setting, (2) preserves model accuracy, and (3) requires no owner cooperation, thereby ensuring robust adversarial defense in resource-constrained TinyML environments.

\subsection{One-sample T-Test}
The one-sample t-test is a parametric hypothesis test used to evaluate whether the mean of a sample significantly differs from a hypothesized population mean \cite{student1908probable}. The test statistic, $t$, defined by Eq. \ref{onesamplet}, measures this deviation relative to the standard deviation estimated from the sample distribution, which serves as an approximation of the population variability. The validity of this test depends on three key assumptions: (1) independence of observations, (2) continuous data, and (3) approximately normal distribution. These assumptions are discussed further in Sec. \ref{sec:runtime}.

\begin{equation}
\label{onesamplet}
t = \frac{\bar{x} - \mu_0}{s/\sqrt{n}}
\end{equation}

The equation for the one-sample t-test statistic is given in Eq. \ref{onesamplet}, where $\bar{x}$ is the sample mean, $\mu_0$ is the hypothesized population mean, $s$ is the sample standard deviation, and $n$ is the sample size. The test statistic, $t$, corresponds to a p-value obtained via Student's t-distribution \cite{student1908probable}. Under the null hypothesis, the sample is assumed to originate from a population with mean $\mu_0$. If the p-value is below a user-defined threshold $P_{th}$, the null hypothesis is rejected in favor of the alternative, indicating that the sample is significantly different from the hypothesized population mean.

\subsection{Power Analysis}
The instantaneous power consumption of a circuit observed through supply voltage fluctuations varies with data-dependent operations \cite{brier2004correlation}. This occurs because the shared power distribution network (PDN), which supplies power to all system components, cannot fully compensate for transient current demands caused by switching activities, resulting in measurable voltage drops \cite{pant2008design}. Prior studies have leveraged such subtle voltage variations to infer sensitive internals, such as cryptographic keys \cite{tehranipoor2023power, wu2025catch}. More broadly, power signature analysis is a method that identifies distinctive features (``signatures'') in power consumption patterns to monitor device operation and has enabled anomaly detection tasks, including program identification \cite{delimitrou2017bolt}, and malware detection \cite{zhang2023trustguard}.

\subsection{Side Channel Analysis for Adversarial Behavior Detection}
Side-channel signals, such as electromagnetic (EM) emissions and power consumption, have gained attention for their potential to detect adversarial activity without requiring access to device internals (i.e., under black-box constraints). {Prior works, such as EMShepherd \cite{ding2023emshepherd} and TrustGuard \cite{zhang2023trustguard}, utilize side-channel leakage captured during inference on FPGA-based accelerator platforms with substantially larger power, memory, and compute budgets than edge microcontrollers to train secondary classifiers in a supervised manner.} For example, EMShepherd leverages EM traces to train an auxiliary classifier to distinguish between benign and adversarial inputs, while TrustGuard uses power traces and trains a multilayer perceptron (MLP) for malware detection. 
% Although effective, these supervised approaches require large volumes of both benign and malicious side-channel traces for training. 
{Additionally, ML-based detection methods \cite{clark2013wattsupdoc, jimenez2019malware} often require large volumes of representative malware training data, which is difficult to obtain and maintain. Because the threat landscape is constantly evolving, any new attack that falls outside the training distribution is at high risk of being misclassified as benign. As a result, such methods may struggle to adapt or scale as adversaries change their tactics, leading to reduced detection performance on previously unseen attacks.}

In contrast, the AdvScan methodology we propose employs statistical hypothesis testing for runtime detection and requires only a small set of benign traces to characterize normal behavior, eliminating the need for supervised training or access to malicious examples. Adversarial inputs are detected based on the irregular neuron activation patterns, which are characteristic of adversarial examples \cite{liu2019abs}{(see Sec. \ref{sec:e1})}. Because detection is based on inherent differences between benign and adversarial activity rather than prior knowledge of specific attack methods, AdvScan can detect adversarial behavior without requiring malicious data for training, thereby overcoming the limitations of prior side-channel-based strategies.

\subsection{Threat Model} \label{sec:four_points}
We consider a black-box TinyML model deployed on a resource-constrained MCU, operating in an untrusted environment. This is representative of a restrictive licensing scenario, where a model owner ships a black-box model as a proprietary compiled binary on a dedicated edge device to an end user with no expectations of further cooperation. The user, as a defender, owns the hardware and interacts with the model only through its normal TinyML interfaces (e.g., inputs and observable outputs) with no access to internal activations, model parameters, or the hardware-specific features. However, the ownership allows the user to instrument the power rail (e.g., by inserting a shunt resistor and sampling the resulting voltage drop with an ADC) to passively monitor power consumption during inference. A practical instance of this setting is an embedded OEM integrator deploying a vendor-supplied software stack that includes a neural network under a restrictive binary/IP license (e.g., a perception module within an Advanced Driver-Assistance System (ADAS) camera pipeline)\cite{eetimes_eyeq5_open, mobileye_valeo_20m_front_camera}. AdvScan uses these side-channel measurements, together with the model’s input–output behavior, to decide whether a given input is likely to be adversarial. In our attack scenario, an adversary is a software-only attacker who aims to compromise model inference integrity by injecting carefully crafted perturbed input aimed to cause misclassification while having no physical or administrative access to the device hardware. They can achieve this by 1) being the model owner who has white-box access to the model parameters or colluding with the model owner to craft tightly tailored AEs; or 2) training a surrogate model using input–output queries to the deployed device and then running standard white-box attacks on that surrogate. A successful defense under this threat model must 1) detect AEs during model inference and 2) identify AEs generated by both known and previously unseen attack algorithms. To assess the efficacy of our AdvScan methodology, we adopt four success metrics established in prior work for evaluating integrity-checking solutions \cite{aramoon2021aid}, which are defined below.
% We consider a black-box TinyML model deployed on a resource-constrained MCU, operating in an untrusted environment. {Under a licensing scenario we consider, a black-box model is shipped as a proprietary compiled binary on a dedicated edge device, and the user, as a defender, can only access inputs and observe outputs, with no access to internal activations, model parameters, or the hardware-specific features.} In our attack scenario, an adversary aims to compromise inference integrity by generating AEs without altering internal model parameters (as described in Sec. \ref{sec:Adv_ex}). A successful defense under this scenario must 1) detect adversarial inputs during model inference and 2) identify AEs generated by both known and previously unseen attack algorithms. To assess the efficacy of our AdvScan methodology, we adopt four success metrics established in prior work for evaluating integrity-checking solutions \cite{aramoon2021aid}, which are defined below.
\begin{enumerate} 
\item \textbf{Effectiveness:} The ability to reliably detect AEs during model inference. 
\item \textbf{Efficiency:} Minimal performance overhead introduced by the detection mechanism, measured by the number of required runtime measurements. 
\item \textbf{Reliability:} Low false-positive (incorrectly flagging benign inputs) and false-negative (failing to detect AEs). 
\item \textbf{Generalizability:} Robust detection performance across different TinyML models, input modalities, {perturbation attacks and user hardware}. 
\end{enumerate}

\section{Motivation \& Problem Formulation} \label{sec:motivation}
TinyML models commonly use aggressive quantization and pruning to fit within limited hardware constraints, making them more vulnerable to adversarial attacks than traditional neural networks (NNs) \cite{preuveneers2023adversarial}. In real-world scenarios, especially black-box deployments such as licensing situations \cite{leroux2022tinymlops}, conventional defenses are not feasible. Therefore, there is a critical need for AE detection techniques that operate in a black-box settings without access to the model's internal parameters.

In this work, we explore power analysis for runtime AE detection under black-box constraints. Power analysis leverages the correlation between an MCU’s power consumption and its data flow \cite{brier2004correlation}. AdvScan is grounded in the hypothesis that \textit{adversarial inputs induce altered neuron activations relative to benign inputs, causing distinct data flow in the NN \cite{liu2019abs} and manifesting in measurable power signature deviations}(see Sec. \ref{sec:e1}). Building on this insight, we propose a power-analysis-based detection approach that remains agnostic to various model architectures, AE generation methods, {data modalities, and user hardware}.

To demonstrate AdvScan’s core assumption, we conducted a motivational experiment {using an STM32F303RC (ARM Cortex-M4) MCU} running a ResNet-20 model from the MLPerf benchmark suite \cite{banbury2021mlperf}, trained on CIFAR-10 dataset \cite{krizhevsky2009learning}. The MCU’s instantaneous power consumption during inference was measured via voltage drops across a $10\Omega$ shunt resistor. For this experiment, we considered two images as inputs classified into the same labels: (1) a benign image, and (2) an AE image generated using the PGD algorithm \cite{mkadry2017towards} with $\epsilon = 0.05$. Initially as shown in Fig. \ref{fig:motivation_traces}(a), raw power traces from these inputs appeared nearly indistinguishable when directly compared, likely due to significant noise inherent in the measurements. The similarity continued when each trace was correlated against a known benign reference trace (i.e., golden template) from the same class (label), resulting in nearly identical Pearson correlation values (see Fig. \ref{fig:motivation_traces}{c}). To mitigate this noise, we collected and averaged 150 power traces each from different inputs of the same classified class for both benign and adversarial scenarios. The averaged power traces, depicted in Fig. \ref{fig:motivation_traces}(b), clearly revealed visual differences between benign and adversarial cases. Further, we generated a golden template by averaging 150 known benign traces from the same class. We then recalculated the Pearson correlation values between this averaged golden template and the averaged benign and adversarial traces, respectively. The resulting gap in correlation values, as demonstrated in Fig. \ref{fig:motivation_traces}(c), provides a preliminary indication that power analysis is a viable method for detecting adversarial perturbations in TinyML while in black-box settings. Despite these promising findings, three primary challenges remain:

\begin{figure}[b]
    \centering
    \includegraphics[width=1\linewidth]{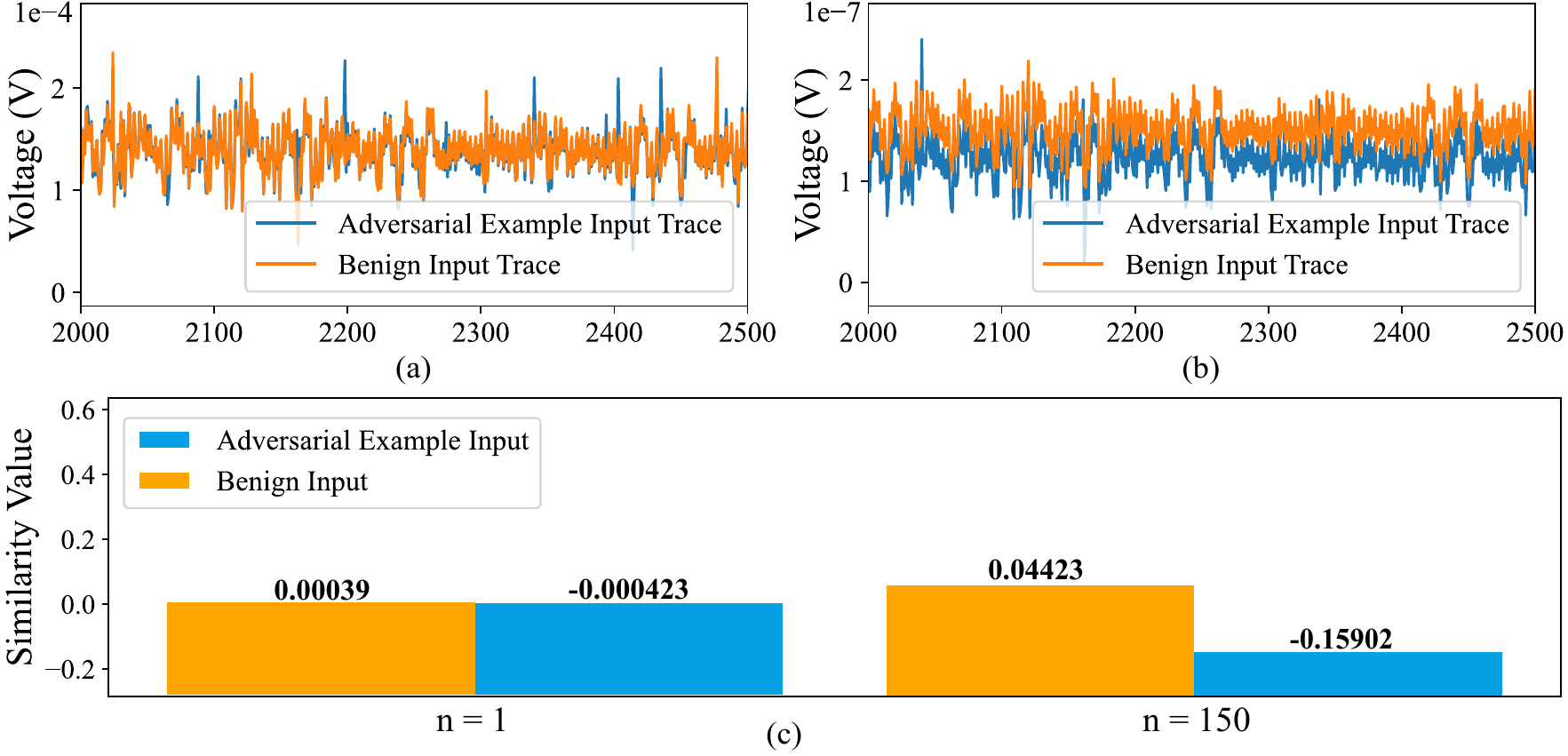}
    \caption{Voltage drop across $10\Omega$ shunt resistor in ResNet-20's final fully connected layer: Overlaid (a) single power traces for benign and AE input, and (b) averaged ($n=150$) power trace for benign and AE input, (c) Pearson correlation between single and averaged benign and AE input traces against a separate class-specific known benign trace.}
    \label{fig:motivation_traces}
\end{figure}
\begin{figure*}[!t]
    \centering
    \includegraphics[width=1\linewidth]{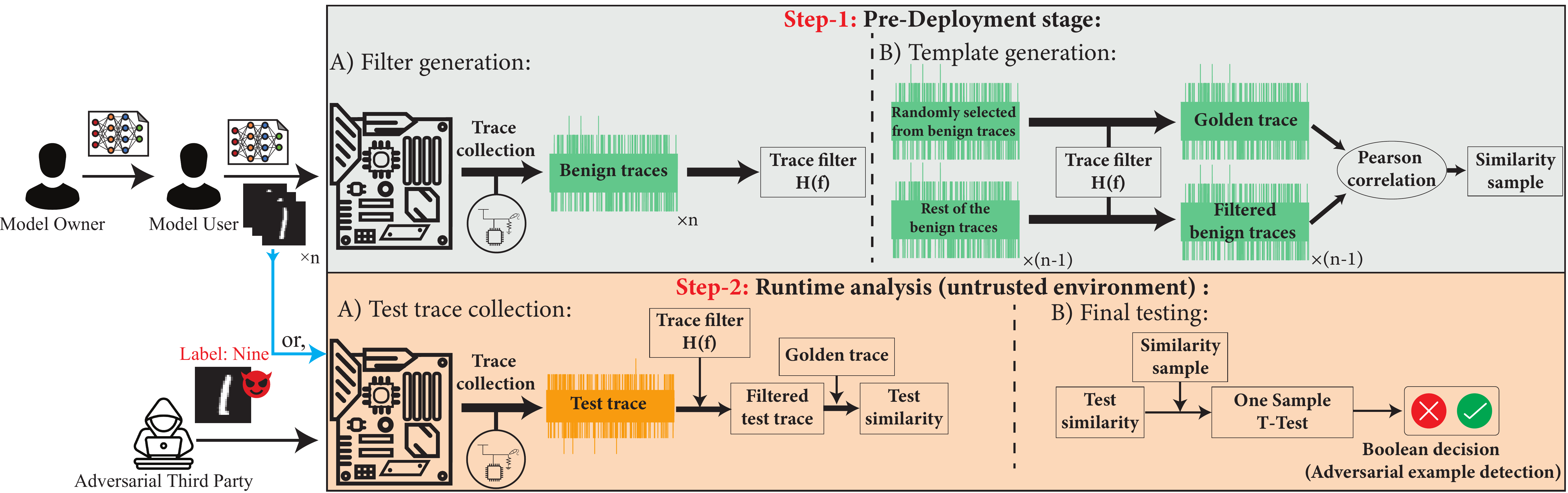}
    \caption{Overview of the proposed AdvScan algorithm that uses power analysis for runtime AE detection.}
    \label{fig:fig2}
\end{figure*}

\begin{enumerate}
\label{sec:prob}

\item \textbf{Noise:} Fig. \ref{fig:motivation_traces}(a) shows that the high noise present in power traces makes AE detection challenging. Typically, the adversary submits a modified input only once, so no averaging can be done to reduce measurement noise. A robust method is needed to detect adversarial behavior from a single trace.

\item \textbf{Template Generation:} Similarity measurements, such as Pearson correlation, provide a mathematically robust algorithm to differentiate between benign and modified traces due to AEs. However, a baseline trace (i.e., a \textit{template} or \textit{signature}) is required for comparison.

\item \textbf{Quantifiable Authentication:} Even two benign traces from the same class can significantly differ due to noise and natural device variations (e.g., thermal effects) (as Fig.  \ref{fig:motivation_traces}(c)). A quantifiable thresholding approach is needed to distinguish benign variation from genuine integrity violations.

\end{enumerate}

\section{AdvScan Algorithm} \label{sec:algo}

To address the challenges outlined in Sec. \ref{sec:motivation}, we propose AdvScan, a power analysis-based runtime AE detection algorithm. AdvScan formulates AE detection as a statistical hypothesis testing problem by constructing a distribution of benign power traces for each model inference class and then applying a one-sample t-test at runtime to identify potential AEs. The process yields a quantifiable measure of confidence (p-value) compared against a user-defined threshold ($P_{th}$), enabling tunable detection sensitivity. AdvScan operates under black-box constraints, requiring no access to the model’s internal architecture or parameters.  Fig. \ref{fig:fig2} overviews AdvScan's two-stage structure: the Pre-Deployment Stage and the Runtime Analysis Stage, each described in detail below.

\subsection{Pre-Deployment Stage}
\label{sec:predeploy}
During the Pre-Deployment Stage of AdvScan, we generate two key outputs for a specific MCU device and model pair: 1) a golden template capturing the expected baseline power consumption and 2) a similarity sample distribution modeling benign input behavior from each class of the considered dataset. To obtain these, we measure the power consumption of the target MCU running inference on ($n$) randomly selected benign inputs (drawn from the model’s training set of the same class). All collected traces are gone through a filtering process (see Sec. \ref{sec:noise}) to reduce the noise. Then, one of these filtered traces is randomly chosen as the golden template, which serves as the baseline for reference. We then correlate each remaining ($n-1$) benign trace with this template to form the similarity sample distribution, characterizing typical variations in power signatures under benign conditions. Finally, at runtime, we collect a test trace for a given test input and correlate it with the corresponding class-specific golden template to obtain a \textit{test similarity value}. A deviation between the test similarity value and the class-specific similarity sample distribution potentially indicates the test input to be an adversarial example (AE). Since power traces are inherently specific to both the hardware and the model, changes to either component necessitate re-executing the Pre-Deployment Stage. The following sections detail the construction processes for the golden template and the similarity sample distribution.

\begin{figure}[t]
    \centering
    \includegraphics[width=1\linewidth]{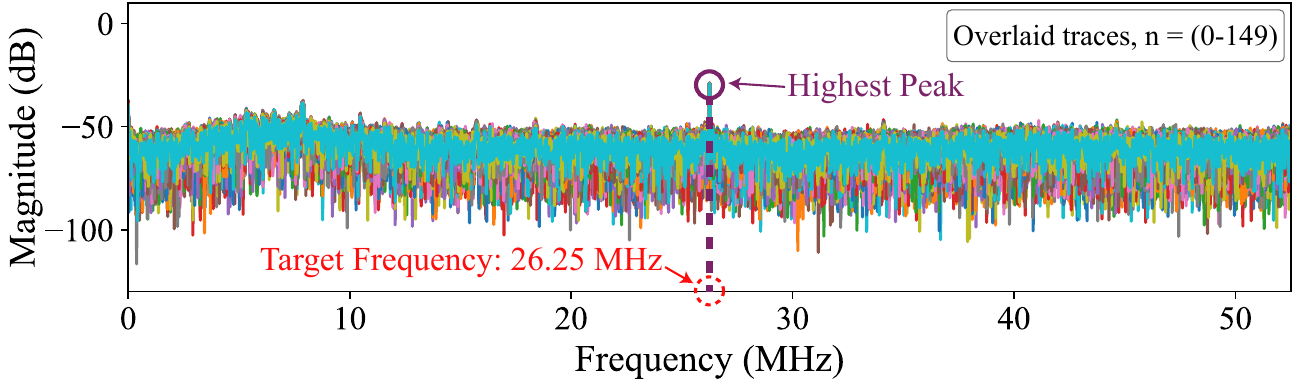}
    \caption{Frequency spectrum analysis of power traces for {STM32F303RC as the target device}. The overlaid Fast Fourier Transform (FFT) of $n=150$ traces show a consistent peak at 26.25 MHz.}
    \label{fig:freq}
\end{figure}

\subsubsection{\textbf{Template Generation}}
\label{sec:noise}
Motivated by the high noise levels highlighted in Sec. \ref{sec:motivation}, we focus on isolating the frequency‑domain components corresponding to computational activity. As reported by Ou et al. \cite{ou2016enhanced}, power/electromagnetic traces typically consist of three elements: a constant DC component, random noise due to environmental or system-level fluctuations, and systematic variations produced by the device’s operations. {In the frequency domain, the DC component is at 0 Hz, and a repeatable peak at 26.25 MHz appears in all $n$ filtered traces collected in Sec. \ref{sec:predeploy} (Fig.~\ref{fig:freq}). Because random noise does not produce such consistent spectral features, we can safely attribute this peak to systematic device activity at the target operating frequency.}

\begin{figure}[t]
  \centering
  \includegraphics[width=\linewidth]{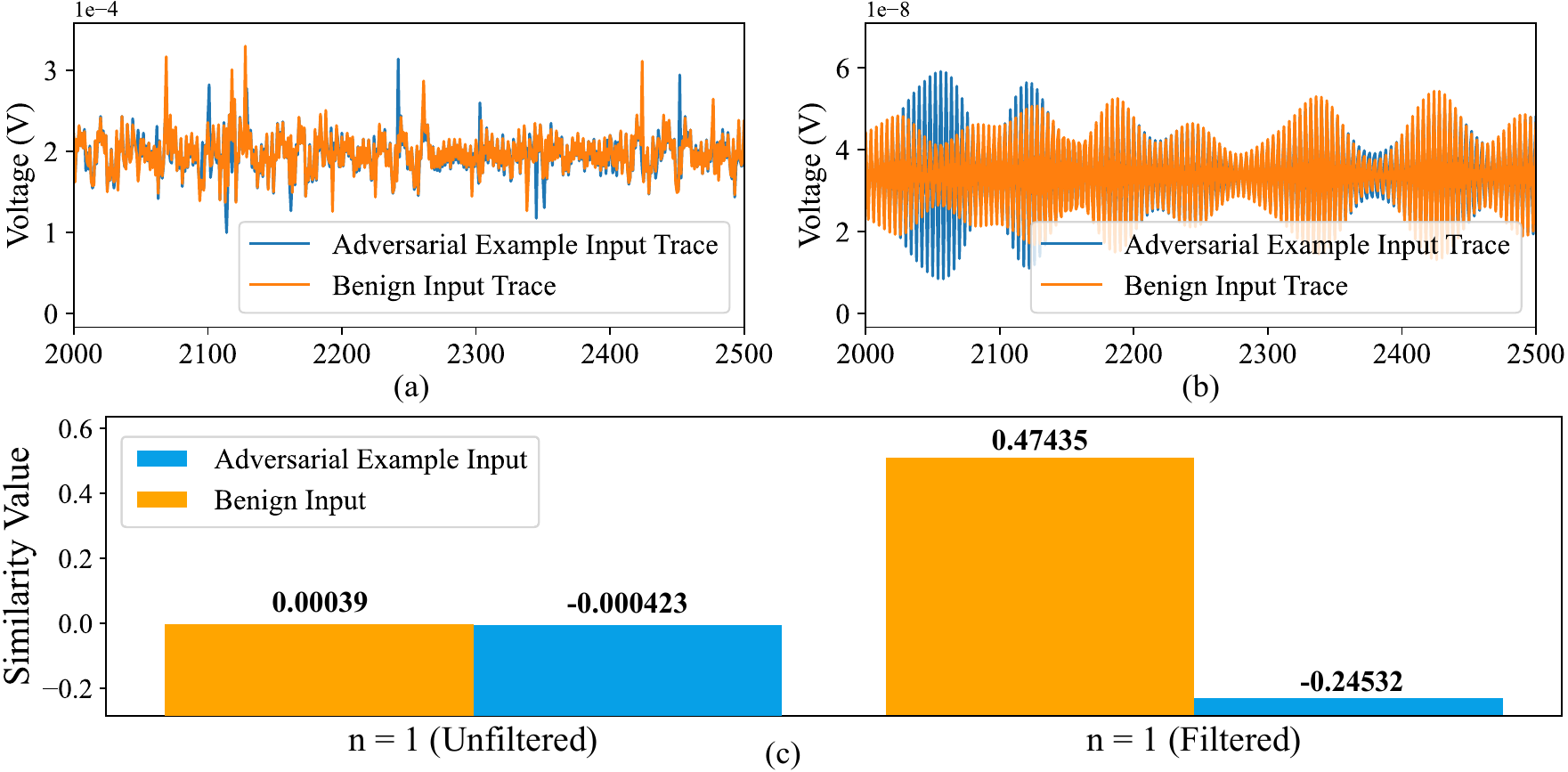}
  \caption{Voltage drop across $10\Omega$ shunt resistor in ResNet-20's final fully connected layer: Overlaid (a) single power traces for benign and AE input, and (b) single filtered power trace for benign and AE input, (c) Pearson correlation between unfiltered and filtered benign and AE input traces to class-specific known benign input trace.}
  \label{fig:motivation_traces_2}
\end{figure}

Based on this observation, we designed a Butterworth band-pass filter centered at 26.25 MHz with a $\pm10$kHz bandwidth and applied it to all $n$ traces. Upon filtering, benign and adversarial traces from the same class diverge visibly when overlaid (Fig. \ref{fig:motivation_traces_2}(b)), and their Pearson correlation with a class-specific, filtered known benign trace exhibits a small gap. This indicates that filtering improves our ability to distinguish adversarial from benign inputs, with only a single power trace. This is crucial since AEs are usually applied only once, allowing only a single power trace to be collected for detection.

\subsubsection{\textbf{Similarity Sample Distribution Generation}}

As shown in Fig. \ref{fig:motivation_traces_2}, although filtering produces clearer distinctions between benign and adversarial traces, the resulting correlation value separation is not perfect. Natural variations, such as system noise, peripheral activity, and temperature fluctuations, contribute to this variability. To address these factors, we use the $n-1$ remaining benign traces from the template generation stage (each from a different input within the same class, excluding the randomly chosen reference trace (i.e., golden template)) as a sample set to represent typical benign variance of power consumption for the hardware. All traces are processed through the same band-pass filter described in Sec. \ref{sec:noise} and correlated with the golden template. {The resulting device- and class-specific empirical distribution of similarity values is referred to as the \textit{similarity sample distribution}. These similarity values can be viewed as samples drawn from an underlying random similarity variable that, in principle, admits a probability density function (PDF). However, AdvScan does not fit or use an explicit parametric PDF; instead, we use only its sample mean and variance in a one-sample t-test} to differentiate benign and anomalous power variation at runtime for AE detection.

\subsection{Runtime Analysis Stage} \label{sec:runtime}
During the runtime analysis stage, AdvScan evaluates each input at inference time to detect potential AEs. The algorithm collects the device’s power consumption (i.e., test trace) while the MCU performs its normal inference on the input. AdvScan then retrieves the golden template and similarity sample distribution associated with $C_{predict}$ from memory. Next, it applies the band-pass filter from Sec. \ref{sec:noise}, and correlates the resulting filtered test trace with the golden template yielding a test similarity value. A one-sample t-test then compares this value against the similarity sample distribution for $C_{predict}$, yielding a probability value (p-value) that reflects how likely it is for the observed trace to arise from the class-specific benign distribution. We hypothesize, building on prior work, that AEs inherently generate abnormal neuron activation patterns \cite{liu2019abs}, which will then alter the MCU’s data flow patterns and thus its power consumption. By modeling each class independently, AdvScan can pinpoint these class-specific deviations and avoid confusion that might arise from aggregating all classes under a single distribution.

AdvScan allows the users to tune detection sensitivity through an adjustable p-value threshold ($P_{th}$) . Lowering $P_{th}$ may increase the false negatives (identifying AEs as benign) whereas a more relaxed threshold does the opposite. In Sec. \ref{sec:all_in}, we demonstrate that with a user-defined threshold of $P_{th}=10^{-5}$, a single power trace is sufficient to identify AEs with very high statistical confidence. The findings strengthen our claim that \textbf{AEs introduce statistically significant shifts in the power signatures of TinyML devices}. To outline the approach, we begin by considering whether the necessary assumptions for a one-sample t-test are satisfied. The one-sample t-test has three key assumptions:

\begin{enumerate}
   \item \textit{Normality Assumption:} The similarity sample distribution should follow a normal distribution. Although Pearson correlation coefficients are bounded between -1 and 1, their empirical distribution can approximate normality when the sample size is sufficiently large.  We verify this using the Shapiro-Wilk test (see Sec. \ref{sec:normality}), which indicates that the similarity sample distribution is highly likely to be normally distributed when a sufficient number of power traces are considered.

   \item \textit{Independence of Observations:} The test similarity value must be independent of the similarity sample distribution. This condition holds because each power trace is derived from an independent inference.

   \item \textit{Continuous Data:} The one-sample t-test assumes continuous data. This condition is met because similarity values, based on Pearson correlation coefficients, are continuous variables ranging from -1 to 1.
\end{enumerate}

Hence, the necessary conditions for using the one-sample t-test are met. Now, we formulate the AE detection problem being considered as a null hypothesis test. Let $\bar{S}$ denote the mean of the benign similarity sample distribution, and $s_{test}$ be the test similarity value for the test input. We define the null hypothesis $(H_0)$ based on a user-defined threshold $(P_{th})$ for the p-value as follows.

\begin{enumerate} 
\item \textbf{Null Hypothesis ($H_0$):} The $s_{test}$ and $\bar{S}$ of the \textit{similarity sample distribution} has no statistically significant difference (i.e., $\text{p-value} \leq P_{th}$).
\item \textbf{Alternative Hypothesis ($H_1$):} The $s_{test}$ and $\bar{S}$ of the \textit{similarity sample distribution} has a statistically significant difference (i.e., $\text{p-value} > P_{th}$).  
\end{enumerate}

If the p-value falls below $P_{th}$, we reject $H_0$ and flag the input as a potential AE. Intuitively, a low p-value indicates that $s_{test}$ is unlikely to belong to the similarity sample distribution. While, in this work, we evaluate the rejection of $H_0$ as an indication of malicious behavior (i.e., AE input), we note that $H_0$ may also be rejected in other scenarios that impact power consumption, such as hardware failures or model modification \cite{paul2025michscan}. Consequently, while AdvScan provides a high-confidence indicator of anomalous behavior, it does not diagnose the exact cause of that anomaly. Nonetheless, in security-critical applications, significant deviations from normal behavior justify a deeper investigation, regardless of the underlying cause, underscoring the value of this detection strategy.

{
\subsection{Runtime Modes and Deployment Configurations}
As shown in Sec. \ref {sec:all_in}, AdvScan can detect AEs using only a single power trace collected during inference on the test input. This capability enables multiple operating modes, where only the trace-collection setup and the execution location differ, while the core algorithms in Secs. \ref{sec:predeploy} and \ref{sec:runtime} remain unchanged. In this study, we consider two concrete runtime modes for AdvScan: (i) a \textbf{gated-output} mode, in which the model’s prediction is released only after the corresponding trace has been verified as benign, and (ii) a \textbf{passive-monitor} mode, in which AdvScan analyzes traces in parallel with normal inference without blocking the output. Both modes can be instantiated under the on-chip and off-chip configurations illustrated in Fig. \ref{fig:setup}.

\subsubsection{Gated-output mode}
In the gated-output mode, AdvScan enforces strict input-integrity checks on the prediction path. For each test input, the TinyML model executes on the MCU while a single power trace is captured for a designated tap layer (see section \ref{sec:layer}). In the on-chip configuration, the MCU samples the shunt-resistor voltage with its internal ADC, stores the resulting trace in on-chip memory, and runs the AdvScan hypothesis test locally before releasing the model’s prediction (Fig. \ref{fig:setup}a). In the off-chip configuration, an external high-speed ADC acquires the trace and streams it to an external MCU, which executes the same AdvScan runtime pipeline and returns a binary accept/reject decision to the TinyML device (Fig. \ref{fig:setup}b). In both cases, the prediction is withheld until the corresponding trace has been accepted as benign, so the detection latency reported in Sec. \ref{sec:efficiency} directly translates into additional end-to-end delay per inference.

\subsubsection{Passive-monitor mode}
In the passive-monitor mode, AdvScan operates as a non-blocking, off-chip watchdog. The TinyML model running on the MCU produces its predictions immediately, while power traces are captured at the shunt resistor and forwarded over a communication protocol (i.e., UART, I$^2$C, or SPI ) to an external MCU that executes the AdvScan detection pipeline asynchronously. The MCU inferencing TinyML itself does not wait for the AE decision and therefore incurs essentially no additional latency on its inference path due to execution of AdvScan; instead, the external MCU raises an alert if the computed $p$-value for a trace falls below the detection threshold $P_{\text{th}}$. This mode thus provides a non-blocking integrity monitor for off-chip deployments, at the cost of allowing a short window in which an AE may be acted upon before being flagged. The communication and processing latencies reported in Sec. \ref{sec:efficiency} characterize how quickly such alerts can be raised in practice.
}

\subsection{Discussion of Merits and Limitations of AdvScan}

We make several observations regarding the merits and limitations of AdvScan:

\subsubsection{Black-Box Operation} AdvScan utilizes power analysis for possible AE detection without requiring model internals or cooperation from model owners. This differs from prior work on AE detection where white-box model access \cite{feinman2017detecting, zhao2023adversarial, lust2020gran, pertigkiozoglou2018detecting} or heavy resource demanding pre-processing \cite{huang2024adversarial, zhu2024pixeldenoise, meng2017magnet} is required.

\subsubsection{Mathematically Quantifiable} AdvScan employs a one-sample t-test, providing a p-value that indicates how likely the test similarity value arises from a distribution built from known benign power traces. Because perturbations alter neuron activation patterns \cite{liu2019abs} and thus the MCU’s power signature (see Sec. \ref{sec:e1}), this offers a statistical framework to identify AEs.

\subsubsection{Configurable Detection} AdvScan enables users to tune detection sensitivity and false-positive rates by adjusting a user-defined threshold $P_{th}$. This flexibility balances between tolerable false alarms (FPs) and missed attacks (FNs) for a given application.
	
\subsubsection{Runtime Assessment} AdvScan requires only a single power measurement ($n=1$) to detect possible AEs, imposing minimal runtime overhead compared to adversarial retraining or input pre-processing-based detection strategies that demand substantial computation and memory requirements.

{
\subsubsection{Generalized AE Detection} AdvScan is configured solely from benign power traces (for both golden templates and similarity distributions), so all possible AEs, regardless of type and generation methodologies, are considered as ``unseen’’ at test time. Since AdvScan’s detection of AE depends only on whether a test trace significantly deviates from this benign similarity distribution, rather than on attack-specific features, the detection methodology is not tied to any particular AE algorithm.
    
\subsubsection{Portability Across Hardware Platforms} AdvScan does not rely on MCU-specific or architecture-specific features, but rather on deviations in power consumption patterns due to altered neuron activations caused by adversarial examples (AEs), it can be applied across different MCU platforms where the power traces are collected across a shunt resistance, as shown in Fig. \ref{fig:setup}. However, deploying AdvScan on new hardware or after substantial model changes will require re-running the Pre-Deployment Stage. 
    
    % \item \textit{Generalized AE Detection:} AdvScan makes no assumptions regarding the MCU, model, data modality, or underlying distribution of power traces. Hence, AdvScan can provide AE detection for arbitrary systems in varied applications.

\subsubsection{Sensitivity to Hardware State and Operating Conditions} AdvScan flags any statistically significant deviation from the benign similarity sample distribution, but it cannot by itself determine whether that deviation was caused by an AE, hardware malfunction, model update, long-term operating-condition drift (e.g., device aging or hardware degradation). In security-critical deployments, any such unexpected deviation from benign behavior is treated as an alert that warrants further investigation. However, false positives caused by operating-condition drift can be resolved by periodically re-running the Pre-Deployment Stage to recalibrate the detection pipeline for the device's current hardware conditions.
}

\begin{table}[b]
    \renewcommand{\arraystretch}{1.3}  % Adds more vertical spacing
    \caption{Overview of TinyML models used for evaluation.}
    \label{table:models}
    \centering
    \begin{tabular}{c c c c}
    \hline
    Use Case & Description & Dataset & Model \\ 
    \hline
    Keyword & Small vocabulary & Speech & \multirow{2}{*}{DS-CNN} \\
    Spotting & (keyword spotting) & Commands & \\
    \hline
    Image & Small image & CIFAR-10 & ResNet-20 \\ \cline{3-4}
    Classification & classification & MNIST & 2 Conv + 1 FC\\
    \hline
    \end{tabular}
\end{table}

\section{Evaluation of AdvScan} \label{sec:setup}
\begin{figure}[t]
    \centering
    \includegraphics[width=1\linewidth]{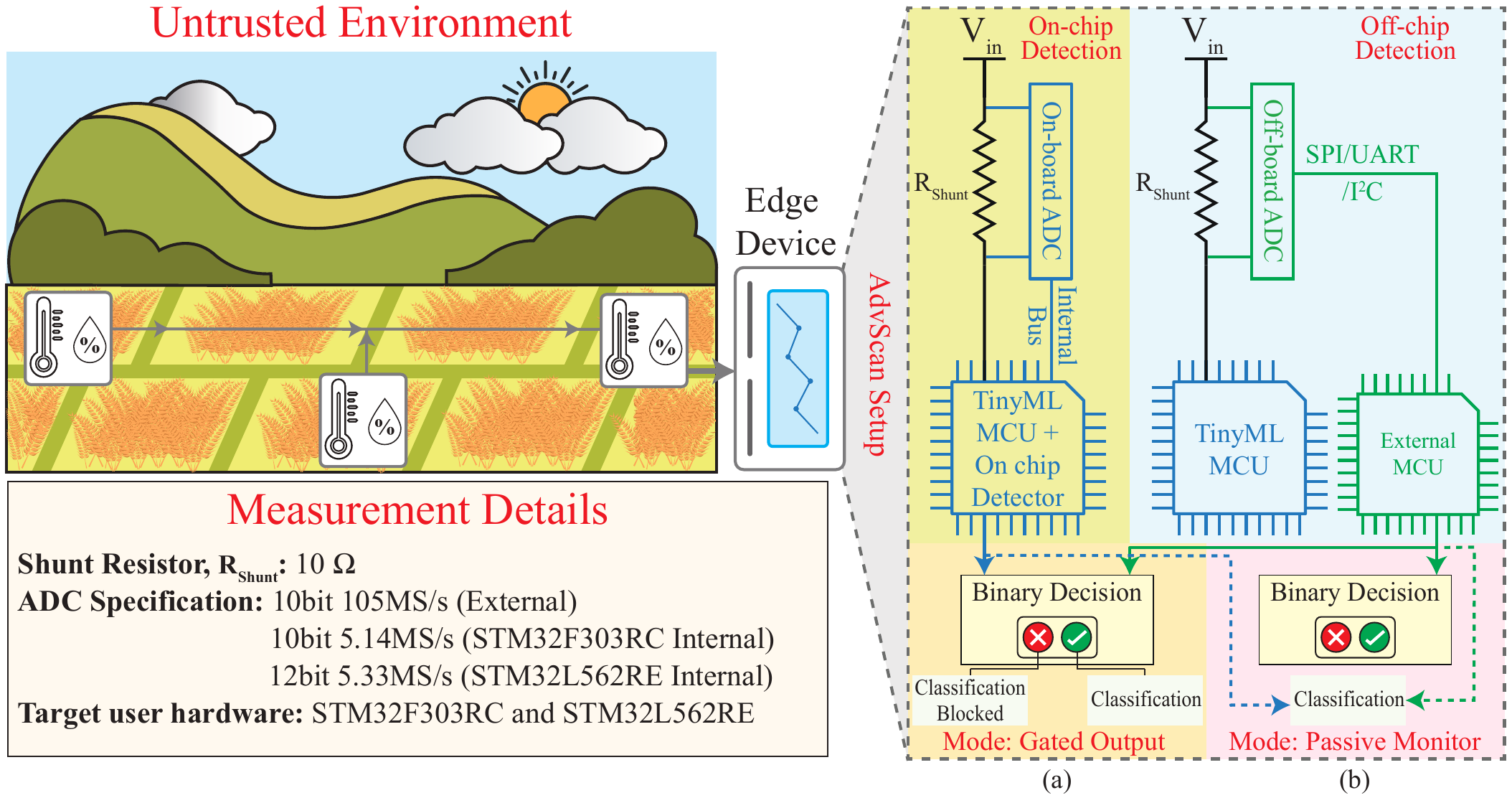}
    \caption{{Measurement setup and operating modes of AdvScan.}}
    \label{fig:setup}
\end{figure}

{
We evaluate the performance of our proposed AdvScan algorithm on two target devices, an STM32F303RC (ARM Cortex-M4) and an STM32L562RE (ARM Cortex-M33), using three TinyML benchmark models from MLPerf-Tiny \cite{leroux2022tinymlops}, as summarized in Table \ref{table:models}. The measurement setup is shown in Fig. \ref{fig:setup}. Power traces are collected by inserting a shunt resistor, $R_{Shunt} = 10\Omega$, in series with the supply rail $V_{in}$ and measuring the resulting voltage drop.

AdvScan supports both on-chip and off-chip configurations, each using a different ADC configuration. For \textit{on-chip detection}, the TinyML MCU utilizes its on-board ADC (10-bit for STM32F303RC, 12-bit for STM32L562RE), configured at its maximum sampling rate (5.14 and 5.33 MS/s, respectively). The shunt voltage is sampled into a buffer of 5,000 samples in on-chip memory via the internal bus, and AdvScan is executed locally on the MCU before the model prediction is released. For \textit{off-chip detection}, we utilize an external 10-bit ADC running at 105 MS/s to capture the shunt voltage, and the resulting traces are transferred to a companion board (i.e., external MCU), which runs AdvScan  and returns a binary accept/reject decision. \change{To keep traces compact and closely tied to model behavior, we target specific layers within each model rather than capturing the entire inference. Here, ``targeting a layer’’ means isolating a recurring segment of the power trace that likely corresponds to that layer’s execution. In this experiment, we use trigger signals to bracket the computation of interest, reducing segmentation and alignment overhead for correlation-based comparisons. These markers provide timing boundaries only and do not expose model internals (e.g., activations, weights), nor are they required by AdvScan. Under a restrictive black-box licensing scenario where modifying/instrumenting the model binary is not permitted, the same segment can instead be located non-invasively within a longer trace using semi-automatic templating/pattern-based alignment as in \cite{trautmann2022semi}.} The resulting modes of operation and their efficiency trade-offs are detailed in Sec. \ref{sec:efficiency}.

}
% {We evaluate the proposed AdvScan algorithm on two target devices, an STM32F303RC MCU (ARM Cortex-M4) and an STM32L562RE (ARM Cortex-M33),} evaluating three TinyML benchmark models from MLPerf-Tiny \cite{leroux2022tinymlops}, as summarized in Table \ref{table:models}. The power traces were collected by measuring the voltage drop across a 10$\Omega$ resistor using a 10-bit ADC sampled at 105 MS/Sec. To maintain concise and manageable trace sizes, we specifically targeted individual layers within each model rather than capturing the entire model execution. This approach leverages the insight that AEs cause anomalous neuron activations throughout the neural network, enabling detection at virtually any model layer (as demonstrated in Section \ref{sec:layer}). To align collected power traces, trigger signals were implemented around each targeted layer's execution. Alternatively, in scenarios where explicit trigger signals cannot be utilized, a semi-automatic templating approach as described in \cite{trautmann2022semi} can be adopted.

\begin{figure*}[!t]
    \centering
    \includegraphics[width=1\linewidth]{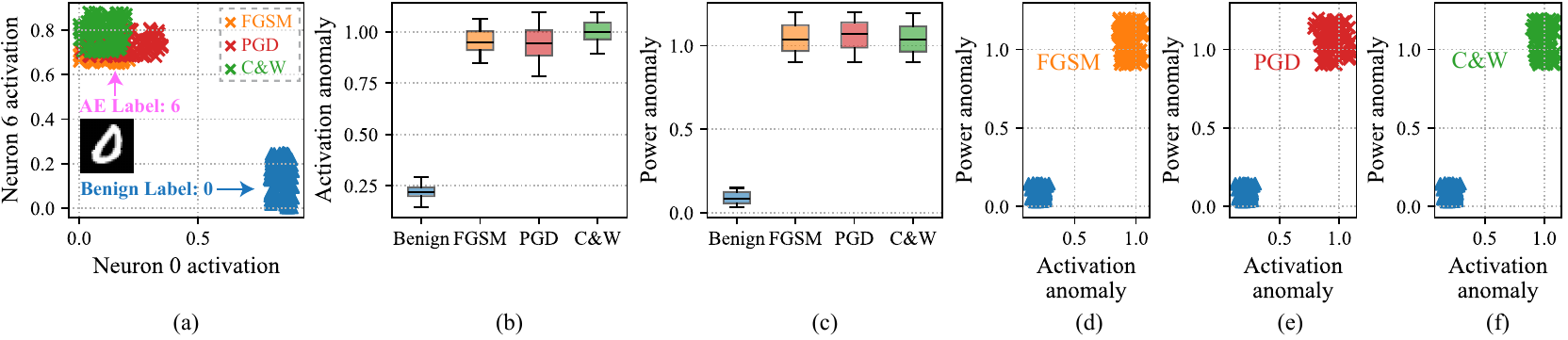}
    \caption{{(a) FC-layer activations of neurons 0 and 6 for benign class-0 inputs and 0$\rightarrow$6 adversarial examples generated by FGSM, PGD, and C\&W. (b) Boxplots of the activation-anomaly score for benign and adversarial inputs. (c) Boxplots of the power-anomaly score for benign and adversarial inputs. (d)–(f) Scatter plots of power-anomaly score versus activation-anomaly score for benign inputs and FGSM, PGD, and C\&W AEs, respectively.}}
    \label{fig:activation}
\end{figure*}

\subsection{Experimental Setup}
To evaluate AdvScan, we implemented the algorithm detailed in Sec. \ref{sec:algo} and provided an overview of its various stages in Fig. \ref{fig:fig2}. The following sections describe each stage of the AdvScan setup in detail.

\subsubsection{\textbf{Benign Trace Collection and Pre-processing}}
During the Pre-Deployment Stage (outlined in Sec. \ref{sec:predeploy}), we collected and preprocessed $n$ power traces from individual benign inputs randomly selected from each target label class for each considered dataset. Further, we identified a target frequency of 26.25 MHz for our MCU (see Sec. \ref{sec:noise}) and designed a Butterworth band-pass filter centered at 26.25 MHz with a $\pm$10 kHz bandwidth. All collected traces were filtered using this band-pass filter. For every target label category, one filtered power trace was randomly chosen as the \textit{golden template}, representing the baseline power signature for benign operation specific to that class. The remaining $(n-1)$ filtered traces for that target class were then correlated with the corresponding \textit{golden template}, forming a class-specific \textit{similarity sample distribution} for benign inputs.

\subsubsection{\textbf{Test Trace Collection and Pre-Processing}}
During the Runtime Detection Stage (see Sec. \ref{sec:runtime}), a single power trace is collected while the model performs inference on a test input, which may be either benign or adversarial. Concurrently, the model predicts a target class for this input. The acquired trace is filtered using the Butterworth band-pass filter (centered at 26.25 MHz, bandwidth $\pm$10 kHz) established during the Pre-Deployment Stage. This filtered test trace is correlated against the golden template corresponding to the predicted class\footnote{While golden templates originate from benign inputs, AEs aim to cause incorrect class predictions. Hence, the analysis uses the golden template and similarity distribution associated with the predicted class.}. Finally, the resulting similarity value is statistically evaluated against the class-specific benign similarity distribution via a one-sample t-test. This test determines if the observed power trace significantly deviates from expected benign behavior to detect AEs.

\subsubsection{\textbf{Adversarial Examples (AE) Generation Methodologies}}
We evaluated AdvScan against three AE generation methodologies: FGSM \cite{goodfellow2014explaining}, PGD \cite{mkadry2017towards}, and C\&W \cite{carlini2017towards}. For FGSM, a single-step gradient-based perturbation was applied, varying the perturbation magnitude ($\epsilon$) from 0.01 to 0.1 in increments of 0.01 \cite{dou2018mathematical}. PGD was performed iteratively within the same $\epsilon$ range, using a step size of 0.01 and projecting the perturbed input back into an $L_{\infty}$-bounded space after each iteration \cite{dou2018mathematical}. For the C\&W attack, we minimized perturbations under various norms ($L_0$, $L_2$, and $L_{\infty}$), systematically varying the optimization constant ($c$) from 0.1 to 1.0 in increments of 0.1 to evaluate its impact. The C\&W optimization objective used in our experiments is detailed in Eq. \ref{cw_ob}.

\subsection{{Root-Cause Analysis of AdvScan's Detection Behavior}}
\label{sec:e1}
To evaluate the underlying assumptions and configuration of AdvScan, we structured our experiments into {3} components:

{
\subsubsection{\textbf{Experiment 1: Activation Anomalies and Power Patterns}} \label{sec:ano}
We first study the effect of adversarial perturbations on internal neuron activations and how this is reflected in power consumption patterns. Our goal is to verify whether an AE can systematically trigger anomalous neuron activations in a TinyML model such that the corresponding power signatures deviate from those of benign inputs. For this evaluation, we consider the STM32F303RC MCU as our target device, and collect $n = 101$ benign power traces from two randomly chosen test inputs from classes 0 and 6 of the MNIST dataset using the TinyML model architecture shown in Table \ref{table:models}. Following Sec. \ref{sec:predeploy}, we construct a class-dependent golden power template for each of these classes: for $class \in \{0,6\}$, we randomly select one benign trace from each class and process it to obtain the golden template $p_{benign}^{(0,6)}$. We then consider targeted AEs generated by FGSM, PGD, and C\&W with a fixed source–target pair (0$\rightarrow$6), so that any observed differences in activations and power traces can be attributed to the perturbations themselves rather than to class-dependent variability. For FGSM and PGD, we set the attack strength to $\epsilon = 0.05$, and for C\&W we use $c = 0.5$; for each attack, we generate 100 AEs. Each AE is applied as a test input to the TinyML model, from which we record the corresponding 10-dimensional FC-layer activation vector $h(x)$ and acquire one FC-layer test power trace $p(x)$.

Fig. \ref{fig:activation}(a) plots the activations of two representative FC neurons corresponding to the true and target classes in this experiment, namely neuron 0 (digit 0) and neuron 6 (digit 6). For benign class-0 inputs, neuron 0 is strongly activated while neuron 6 remains low, whereas successful 0$\rightarrow$6 AEs suppress neuron 0 and strongly activate neuron 6. As shown in the figure, benign samples (blue) form a compact cluster at high neuron-0 and low neuron-6 activation, while AEs generated by FGSM, PGD, and C\&W occupy distinct regions with substantially increased neuron-6 activation. This illustrates that, even within the same true class, adversarial perturbations drive the FC-layer activations away from the benign manifold.

Next, to quantify the deviations in the full 10-dimensional activation space for the FC-layer and relate them to the observed power traces, we define two anomaly scores. Let $\mu_{benign}$ denote the mean FC-layer activation vector over benign class-0 inputs. For any test input x, we define the activation-anomaly score as $D_{act}(x) = \parallel h(x) - \mu_{benign}\parallel_{2}$, which measures how far its activation pattern departs from typical benign behavior. In parallel, $p_{benign}^{(0)}$ and $p_{benign}^{(6)}$ denote the FC-layer golden power templates for classes 0 and 6, respectively. For a benign class-0 input x, we compute a similarity value $S(x)$ between $p(x)$ and $p_{benign}^{(0)}$. For a 0$\rightarrow$6 AE, which is misclassified as class 6, we instead compute $S(x)$ between $p(x)$ and $p_{benign}^{(6)}$, i.e., the template of the target/predicted class. We then define the power-anomaly score, $D_{pow} (x) = 1 - S(x)$, which is small when the measured power trace closely matches the corresponding benign template and increases as the trace deviates from it.

Fig. \ref{fig:activation} (b) and (c) summarize these anomaly scores for benign class-0 inputs and for FGSM, PGD, and C\&W AEs. The boxplots show a clear separation between benign and adversarial cases in both activation-anomaly and power-anomaly scores. Next, Fig. \ref{fig:activation}(d)–(f), plot the power-anomaly score versus the activation-anomaly score for benign inputs (blue) and for FGSM, PGD, and C\&W AEs, respectively. In each case, benign samples cluster near the origin, corresponding to low activation and power anomaly, whereas the adversarial examples form a distinct cluster in the high-anomaly region. The clear separation between these clusters shows that the same inputs that are activation-anomalous are also power-anomalous, suggesting a correlation between internal activation changes and power-consumption patterns for the considered TinyML layer.
}

\begin{figure*}[!t]
    \centering
    \includegraphics[width=1\linewidth]{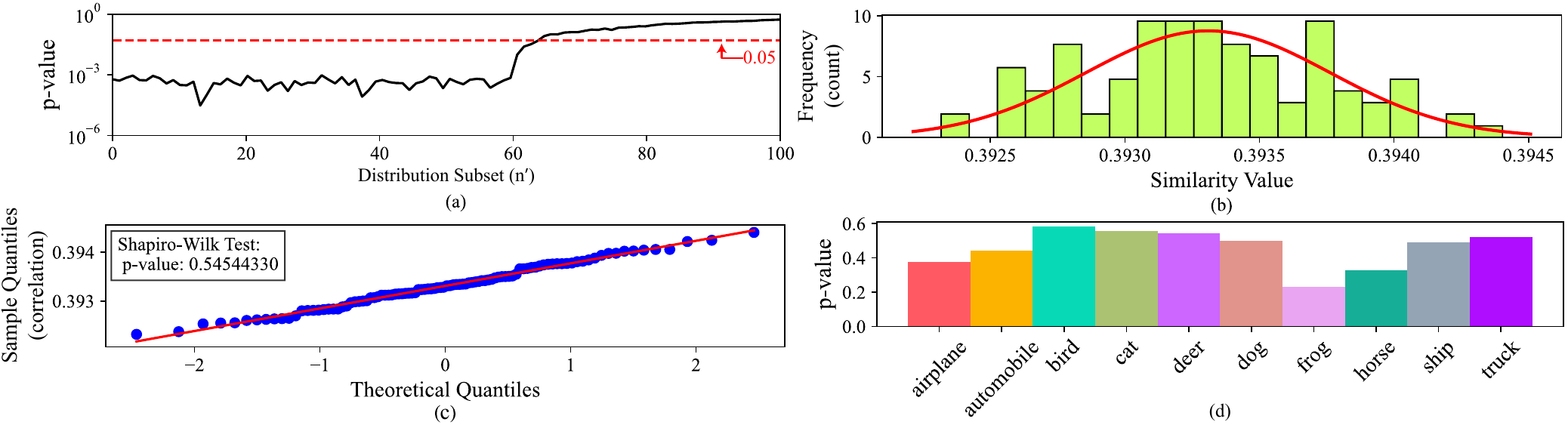}
    \caption{(a) Evaluation of normality of similarity sample distribution using a Shapiro-Wilk Test for varying sample sizes ($n'$). (b) Distribution of similarity values showing a histogram with the fitted normal curve. (c) Q-Q plot with Shapiro-Wilk test results (p-value = 0.5454) confirming normality, (d)  Shapiro-Wilk test (p-values) across multiple CIFAR-10 classes confirming normality consistently holds (n=100) for each class-specific similarity distribution. All traces are collected from the fully connected layer of ResNet-20 model.}
    \label{fig:fig_4}
\end{figure*}

\begin{figure}[t]
    \centering
    \includegraphics[width=1\linewidth]{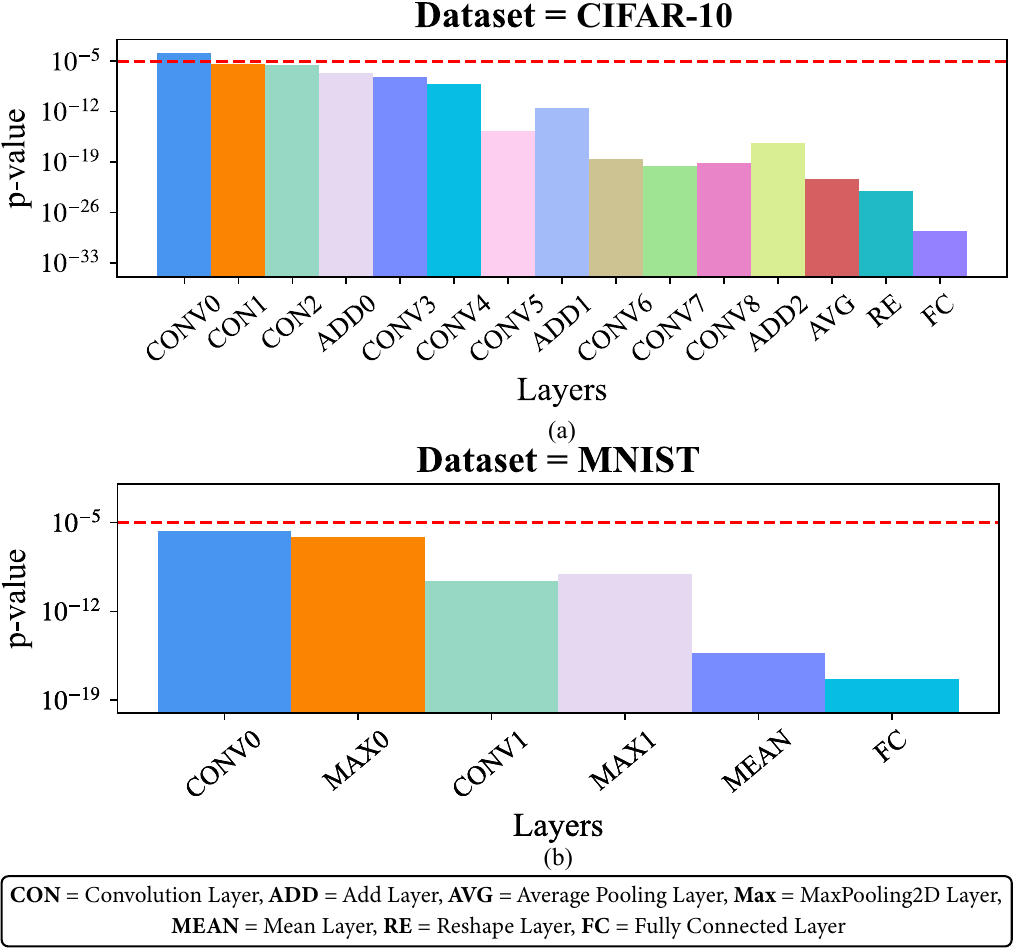}
    \caption{Layer-wise p-value for AE detection}
    \label{fig:fig6}
\end{figure}

\begin{figure*}[!t] 
    \centering
    \includegraphics[width=1\linewidth]{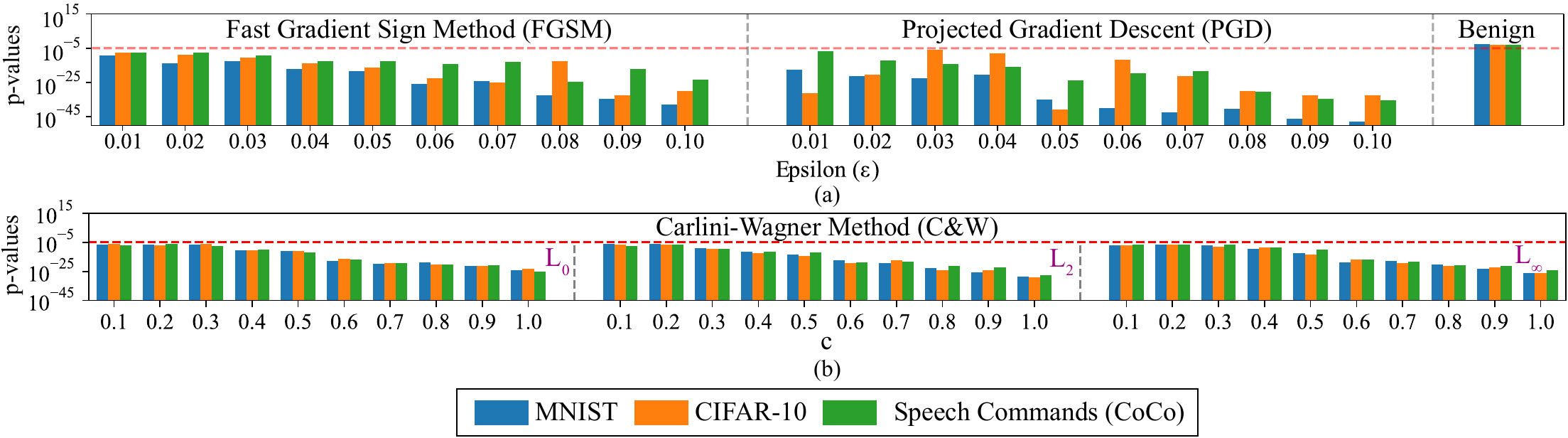}
    \caption{Runtime AE detection using AdvScan {on the target device, an STM32F303RC microcontroller (MCU),} across multiple model architectures (see Tbl.~\ref{table:models}). For each model, the reported p-values quantify the statistical deviation of power-consumption traces from benign baselines for AEs generated using FGSM, PGD, and C\&W attacks constrained under $L_0$, $L_2$, and $L_{\infty}$ norms. The detection threshold ($P_{th} = 10^{-5}$; dashed red line) illustrates AdvScan’s ability to separate benign inputs (typically high p-values) from AEs (low p-values).}
    \label{fig:fig_7}
\end{figure*}

\begin{figure}[t]
    \centering
    \includegraphics[width=1\linewidth]{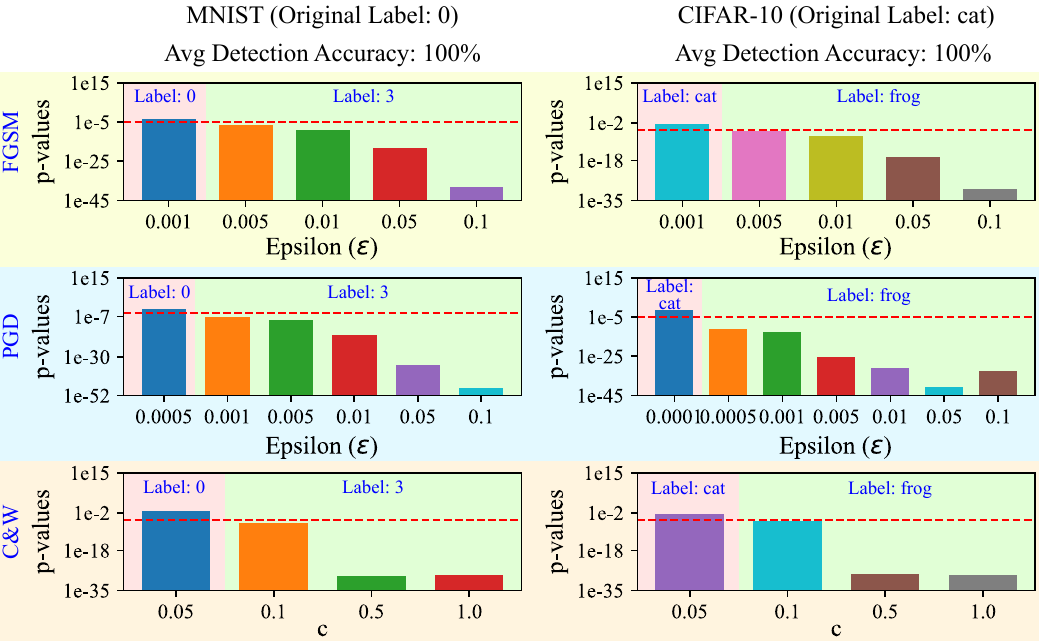}
    \caption{{Evaluation of AdvScan’s effectiveness against an attacker, with knowledge of AdvScan’s detection threshold and pipeline, and iteratively reduces perturbation severity until the AE is no longer flagged.}}
    \label{fig:fig8}
\end{figure}

\subsubsection{\textbf{Experiment 2: Verification of Necessary Assumptions}} \label{sec:normality}
{Next,} we evaluate whether the benign similarity sample distribution satisfies the normality assumption required by the one-sample t-test. It is important to note that the remaining assumptions for the t-test, specifically the independence of observations and the continuity of the data, are satisfied by the AdvScan algorithm and design choices made. Each power trace is collected independently from separate input inferences, and the similarity values, computed as correlation coefficients, are continuous and naturally bounded between -1 and 1. Thus, we focused our verification efforts on normality.

To verify the normality assumption required for the one-sample t-test, we applied the Shapiro-Wilk test to the similarity sample distribution obtained from the STM32F303RC MCU measurements. The Shapiro-Wilk test assesses how closely the observed data aligns with a theoretical normal distribution, yielding a p-value as an indicator of normality. Specifically, a p-value below 0.05 implies a statistically significant deviation from normality, whereas values above 0.05 indicate no substantial departure from a normal distribution. In our evaluation, we initially gathered $n = 101$ power traces for a ResNet-20 model trained on CIFAR-10 (label class: "deer"), from which we randomly selected a single golden trace. Following the filtering and correlation procedures described in Sec. \ref{sec:predeploy}, we obtained 100 similarity values forming the benign similarity distribution. We then tested subsets ($n'$) of these values with the Shapiro-Wilk test to determine the minimal sample size that reliably reflects normality. As shown in Fig. \ref{fig:fig_4}(a), the resulting p-values notably increase and surpass the 0.05 threshold for subsets with $n' \geq 60$, highlighting that larger subsets exhibit clearer normal distribution characteristics. Additionally, Fig. \ref{fig:fig_4}(b) displays a histogram of the entire $n = 100$ sample overlaid with a fitted Gaussian curve, showing symmetry around the mean, consistent with normality. The corresponding Q-Q plot in Fig. \ref{fig:fig_4}(c) further confirms normality, with data points closely following the diagonal reference line. Lastly, Fig. \ref{fig:fig_4}(d) summarizes Shapiro-Wilk test results across all CIFAR-10 classes, indicating consistently high p-values that support the normality assumption for all class-specific distributions.

\subsubsection{\textbf{Experiment 3: AdvScan Target Layer Selection}} \label{sec:layer}
For a TinyML model considering AE inputs, anomalous neuron activations typically accumulate and become more pronounced as data propagates through deeper layers. Consequently, power consumption captured at a single, strategically chosen layer can effectively reflect these disturbances, including those originating from earlier layers. By analyzing only the most sensitive layer instead of the entire network, we reduce the data size and computational overhead without compromising AE detection reliability. To identify the optimal layer for detection, we analyzed power consumption signatures across all layers of two TinyML models: a ResNet-20 trained on CIFAR-10 and a simpler architecture (2 Conv + 1 FC) trained on MNIST.

For this experiment, we first {considered the STM32F303RC MCU as our target device and} collected a power trace at each candidate layer during model inference on a representative AE generated via the PGD method with perturbation magnitude $\epsilon=0.05$ over 40 iterations. Each trace was filtered and correlated against its class-specific golden template corresponding to the model's inference for the AE.  The resulting correlation (similarity) value was then statistically compared to the corresponding benign similarity distribution using a one-sample t-test, yielding a p-value indicative of adversarial perturbations. We repeated this procedure systematically for each layer of both models using the same adversarial input to ensure consistency. Figure \ref{fig:fig6} illustrates the resulting p-values for each targeted layer within the two models.

Our results show a clear trend where deeper layers consistently yield lower p-values, signifying greater sensitivity and stronger response to adversarial perturbations. Specifically, earlier layers demonstrated limited distinguishability between benign and AE inputs, while the final fully connected layer consistently produced the lowest p-values across both model architectures. This indicates that the final layer is most sensitive to adversarial behavior, thus making it a favorable candidate for reliable runtime detection. Based on these observations, we selected the final fully connected layer as the primary target for subsequent experiments.

\subsection{Evaluation of the AdvScan Algorithm}
\label{sec:all_in}
Next, we evaluated AdvScan's AE detection capability across various experimental scenarios. Specifically, we evaluate AdvScan against benign and adversarial inputs generated by three prominent attack methods: FGSM, PGD (using the $L_{\infty}$ norm), and C\&W (applied with $L_0, L_2, \text{and} \ L_{\infty}$ norms) across three model architectures (see Sec. \ref{table:models}), yielding 18 total scenarios. For each inference, we collected a single power trace at the model's final fully connected (FC) layer, filtered this trace, and correlated it with the prediction-specific golden template. The resulting test similarity value was then statistically compared to the corresponding benign similarity distribution with a one-sample t-test. To systematically evaluate AdvScan’s robustness and utility, we measured performance across four success metrics (effectiveness, efficiency, reliability, and generalizability) as defined in Sec. \ref{sec:four_points} .

\subsubsection{\bf{Success Criterion 1  (Effectiveness)}}

The effectiveness of AdvScan refers to its capability to accurately distinguish AEs from benign inputs at runtime. Figure \ref{fig:fig_7} depicts the p-values obtained during the runtime detection phase, considering test inputs given as benign and AEs generated by FGSM, PGD, and C\&W attacks under multiple perturbation norms. For this experiment, we considered the STM32F303RC MCU as our target device, and a user-defined detection sensitivity threshold $P_{th} = 10^{-5}$, which is shown as a dashed red line in Figure \ref{fig:fig_7}. AdvScan's runtime analysis stage flags a test input as a potential AE when the obtained p-value from the one-sample t-test falls below this threshold. This indicates the test similarity has significantly deviated from the mean of the benign similarity distribution.

\change{Next, we continued our evaluation of AdvScan’s effectiveness by considering an attack scenario in which an attacker is aware of AdvScan's detection pipeline and has white-box access to its internal AE detection probabilities, including the detection threshold $P_{th}$, and uses this knowledge to launch a detector-aware threshold-guided evasion attack}. Figure \ref{fig:fig8} summarizes this experiment: for a representative benign input from MNIST and CIFAR-10, and for each attack (FGSM, PGD, and C\&W), the attacker starts from a relatively large perturbation budget and then evaluates a sequence of decreasing perturbation strengths. \change{For FGSM and PGD, attack budget is , $\epsilon \in \{0.50, 0.10, 0.05, 0.01, 0.005, 0.001, \ldots\}$), and for C\&W, the attack budget is, $c \in \{5.0, 1.0, 0.5, 0.1, 0.05, 0.01, \ldots\}$).} We continue decreasing the perturbation strength until AdvScan no longer flags the input as an AE. We then query the model with this final input and record whether the prediction matches the ground-truth label. We repeat this adaptive procedure $n = 20500$ times for each attack. Different trials terminate at different perturbation levels due to the randomized perturbation process. However, across all $n = 500$ runs, we do not observe any instance in which the final input simultaneously (i) causes misclassification and (ii) yields a p-value $\geq P_{th}$. \change{These results suggest that, for the set of trials evaluated, no successful misclassification was observed to evade detection under a threshold-guided strategy.}

\change{
AdvScan’s runtime decision is class-conditioned. For each test input, we correlate the measured (filtered) power trace with the golden template of the model’s predicted class and compare the resulting similarity value against the benign similarity distribution of that same predicted class via a one-sample t-test. Thus, for an attack to both succeed (cause misclassification) and evade detection, it must produce an input whose power-trace similarity falls within the benign range for the (misclassified) predicted class (i.e., yields a p-value $> P_{th}$). Based on this experiment, while misclassification is maintained for a given input, the resulting traces remain statistically inconsistent with the benign distribution of the predicted class, so the similarity statistic lies in the tail, and the p-value remains below $P_{th}$ (see Fig. \ref{fig:fig_7}). Consequently, in this threshold-guided search, the first point at which the input is no longer flagged coincides with the perturbation becoming too small to cause misclassification, at which point the input reverts to benign behavior, and therefore is not flagged by AdvScan (see Fig. \ref{fig:fig8}).}

{

}

\subsubsection{\bf Success Criterion 2 (Efficiency)} \label{sec:efficiency}

{
Efficiency measures AdvScan’s ability to achieve reliable AE detection with minimal runtime overhead. For each inference, AdvScan requires only a single power trace measurement, enabling detection in a reasonable scenario where an AE may be applied to the model only once. We evaluate the efficiency of AdvScan along two axes: (1) detection quality and (2) overall latency for both off-chip and on-chip deployment.

\textbf{\textit{Detection Quality:}} Table \ref{table:acc} summarizes AdvScan’s detection performance across the three TinyML benchmarks (MNIST, CIFAR-10, and Speech Commands), multiple model architectures, and two target MCU platforms. Using a user-defined detection threshold of $P_{th} = 10^{-5}$, AdvScan correctly classifies 318,360 out of 318,400 test cases, comprising both benign inputs and AEs generated via FGSM, PGD-$L_{\infty}$, and C\&W attacks under $L_0$, $L_2$, and $L_{\infty}$ norms. This corresponds to an overall detection accuracy of $ \approx 99.987\%$, with only 40 test cases misclassified. 

\textbf{\textit{Overall Latency:}} \change{To evaluate the overall detection latency of the AdvScan algorithm, we define end-to-end detection time as $t_{e2e} \approx \max \left( t_{inf}, t_{cap} \right) + t_{tx} + t_{proc}$, where $t_{inf}$ is the baseline inference time of the user model, $t_{cap}$ is the trace capture time, $t_{tx}$ is the communication time to transfer the trace to the analysis point (if any), and $t_{proc}$ is the AdvScan processing time.} In our implementation, each ADC sample (10-bit on STM32F303RC and 12-bit on STM32L562RE and 10 bit on external ADC) is stored in a 16-bit word for alignment, so a 5,000 sample trace corresponds to $S = \frac{16 \times 5000}{8} = 10,000$ bytes in transit with the experimental setup as in Sec. \ref{sec:setup}.

In the off-chip detection configuration, power traces are acquired using an external 10-bit ADC and transferred to an external MCU for analysis over standard communication interfaces, e.g., UART, I$^2$C, and SPI (see Fig. \ref{fig:setup} and Sec. \ref{sec:setup}). In this setup, the user model requires $t_{inf} \approx 0.9631$ ms per test input. The ADC operates at 105 MS/s, resulting in a capture time of $t_{cap}^{off}=\frac{5000}{105\times10^{6}} \approx 0.0476$ ms for a 5,000 sample trace. AdvScan’s processing pipeline further requires, on average, $t_{proc}\approx1.65$ ms. \change{Here, we empirically measure the on-device latency components $t_{inf}$, $t_{cap}$, and $t_{proc}$ using hardware timers. For off-chip configurations, transfer latency is estimated analytically from trace size and interface throughput. This is because direct end-to-end measurement is not currently instrumented in our experimental testbed.} Table \ref{tab:latency} reports the end-to-end detection latency for off-chip analysis when transmitting and processing a single 10 kB trace: $t_{det}=870.668$ ms over UART (115.2 kbit/s), $227.612$ ms over I$^2$C (400 kbit/s), and $12.612$ ms over SPI (8 Mbit/s).

\change{
For the on-chip detection configuration, the STM32F303RC and STM32L562RE use their on-chip ADCs (10-bit at 5.14 MS/s and 12-bit at 5.33 MS/s, respectively) to capture 5,000-sample traces directly into on-chip memory. This results in a capture time of $\approx 1$ms for both target devices. Because the on-chip detection configuration requires no trace transfer, we exclude transfer latency from this calculation. This yields an end-to-end on-chip detection latency of approximately 2.6 ms on both MCUs. In our setup, the user model requires $t_{inf} \approx 0.9631$ ms per input, implying an additional latency of $\approx 1.64$ ms beyond baseline inference. Table \ref{tab:comparison} summarizes how AdvScan compares to prior runtime-based AE detection approaches in terms of deployability, assumptions, and measured overhead. Prior methods typically rely on trace transfer and/or off-chip processing and may require adversarial or malicious data during training, which reduces practicality for constrained edge deployments. In contrast, AdvScan achieves 99.984\% detection accuracy while maintaining a 2.623 ms end-to-end on-chip detection latency on both MCUs and meeting the stated deployment requirements in Sec. \ref{sec:four_points}.
}

\begin{table}[t]
\centering
% \color{blue}
\caption{End-to-end alert latency for a single 10kb trace.}
\label{tab:latency}
\resizebox{\columnwidth}{!}{%
\begin{tabular}{@{}ccccc@{}}
\toprule
\multirow{2}{*}{\begin{tabular}[c]{@{}c@{}}Communication\\ Protocol\end{tabular}} & \multirow{2}{*}{\begin{tabular}[c]{@{}c@{}}Transfer\\ Rate $^*$\end{tabular}} & \multirow{2}{*}{\begin{tabular}[c]{@{}c@{}}Total latency\\ (Off-chip)\end{tabular}} & \multicolumn{2}{c}{\begin{tabular}[c]{@{}c@{}}Total latency\\ (On-chip)\end{tabular}} \\ \cmidrule(l){4-5} 
 &  &  & STM32F303RC & STM32L562RE \\ \midrule
UART & 115.2 kbit/s & 870.668 ms & \multirow{3}{*}{2.623 ms} & \multirow{3}{*}{2.588 ms} \\ \cmidrule(r){1-3}
I\textsuperscript{2}C & 400 kbit/s & 227.612 ms &  &  \\ \cmidrule(r){1-3}
SPI & 8 Mbit/s & 12.612 ms &  &  \\ \bottomrule
\end{tabular}%
}
\begin{flushleft}
\footnotesize ${}^*$ Transfer times assume per-byte framing: UART 8N1 $\Rightarrow$ 10 bits/byte; I\textsuperscript{2}C (byte + Acknowledgement) $\Rightarrow$ 9 bits/byte; SPI $\Rightarrow$ 8 bits/byte.
\end{flushleft}

\end{table}

\newcommand{\rothead}[1]{\rotatebox{90}{\parbox{2.4cm}{\centering\bfseries #1}}}
\newcommand{\rot}[1]{\rotatebox{90}{{#1}}}

\begin{table}[b]
\centering
\renewcommand{\arraystretch}{1.25}
\setlength{\tabcolsep}{5pt}

\caption{\change{Comparison between AdvScan and prior runtime side channel AE detection methodologies}}
\label{tab:comparison}
\resizebox{\columnwidth}{!}{%
% {\color{blue}%
\begin{tabular}{|c|c|c|c|c|c|c|c|c|c|}
\hline
\multicolumn{2}{|c|}{} &
\rothead{\begin{tabular}[c]{@{}c@{}}Edge\\ Deployment\end{tabular}} &
\rothead{\begin{tabular}[c]{@{}c@{}}Malicious data \\ Requirements\\ for Training\end{tabular}} &
\rothead{\begin{tabular}[c]{@{}c@{}}Hardware\\ Agnostic\end{tabular}} &
\rothead{\begin{tabular}[c]{@{}c@{}}Attack\\ Agnostic\end{tabular}} &
\rothead{\begin{tabular}[c]{@{}c@{}}Runtime\\ Detection\end{tabular}} &
\rothead{\begin{tabular}[c]{@{}c@{}}On chip\\ Detection\\ Capability\end{tabular}} &
\rothead{\begin{tabular}[c]{@{}c@{}}Detection\\ Accuracy\end{tabular}} &
\rothead{\begin{tabular}[c]{@{}c@{}}Detection\\ Latency\end{tabular}} \\ \hline

\multirow{4}{*}{\rot{Prior works}} & \cite{feinman2017detecting} &
\cellcolor{red!25}$\times$ &
\cellcolor{red!25}$\checkmark$ &
\cellcolor{green!25}$\checkmark$ &
\cellcolor{red!25}$\times$ &
\cellcolor{green!25}$\checkmark$ &
\cellcolor{red!25}$\times$ &
92.6\% & -- \\ \cline{2-10}

& \cite{wang2022manda} &
\cellcolor{red!25}$\times$ &
\cellcolor{red!25}$\checkmark$ &
\cellcolor{green!25}$\checkmark$ &
\cellcolor{green!25}$\checkmark$ &
\cellcolor{green!25}$\checkmark$ &
\cellcolor{red!25}$\times$ &
98.41\% & 0.26ms \\ \cline{2-10}

& \cite{ding2023emshepherd} &
\cellcolor{red!25}$\times$ &
\cellcolor{green!25}$\times$ &
\cellcolor{green!25}$\checkmark$ &
\cellcolor{green!25}$\checkmark$ &
\cellcolor{green!25}$\checkmark$ &
\cellcolor{red!25}$\times$ &
$>90\%$ & 169ms \\ \cline{2-10}

& \cite{alam2024advhunter} &
\cellcolor{red!25}$\times$ &
\cellcolor{green!25}$\times$ &
\cellcolor{red!25}$\times$ &
\cellcolor{green!25}$\checkmark$ &
\cellcolor{green!25}$\checkmark$ &
\cellcolor{red!25}$\times$ &
98.74\% & -- \\ \hline

\multicolumn{2}{|c|}{Our Work} &
\cellcolor{green!25}$\checkmark$ &
\cellcolor{green!25}$\times$ &
\cellcolor{green!25}$\checkmark$ &
\cellcolor{green!25}$\checkmark$ &
\cellcolor{green!25}$\checkmark$ &
\cellcolor{green!25}$\checkmark$ &
99.984\% & 2.623ms \\ \hline
\end{tabular}%
}
% }
\end{table}

From a detection perspective, these results show that even in the worst-case off-chip configuration (using a 115.2 kbit/s UART link), AdvScan can raise an alert in under one second, while an on-chip deployment enables near-real-time detection with only $\approx 2.6$ ms of additional delay per inference. In practice, these latencies support two deployment modes. In a gated-output configuration, the device withholds the model’s prediction until AdvScan has verified the corresponding trace as benign, so the detection times in Table \ref{tab:latency} (and the 2.6 ms on-chip delay) directly correspond to the extra end-to-end latency before any action is taken. Alternatively, in extremely high-throughput scenarios where any additional delay on the inference path is unacceptable, AdvScan can operate as a passive monitor that analyses side-channel traces in parallel with normal inference, introducing minimal overhead on inference latency but possibly allowing malicious behaviour to occur briefly before a flag is triggered. This flexibility makes AdvScan suitable for deployment on resource-constrained edge platforms in both off and on-chip monitoring.
}

% \begin{table}[t]
% \centering
% \color{blue}
% \caption{End-to-end alert latency for a single 10kb trace.}
% \label{tab:latency}
% \resizebox{\columnwidth}{!}{%
% \begin{tabular}{@{}ccccc@{}}
% \cmidrule(r){1-3} \cmidrule(l){5-5}
% \textbf{\begin{tabular}[c]{@{}c@{}}Communication\\ Protocol\end{tabular}} & \textbf{\begin{tabular}[c]{@{}c@{}}Transfer\\ Rate $^*$\end{tabular}} & \textbf{\begin{tabular}[c]{@{}c@{}}Total latency \\ (\boldmath$(t_{\mathrm{det}} = t_{\mathrm{cap}} + t_{\mathrm{tx}} + t_{\mathrm{proc}})$)\\ (Off-board)\end{tabular}} & \textbf{} & \textbf{\begin{tabular}[c]{@{}c@{}}Total latency \\ (\boldmath$(t_{\mathrm{det}} = t_{\mathrm{cap}} + t_{\mathrm{proc}})$)\\ (On-board)\end{tabular}} \\ \cmidrule(r){1-3} \cmidrule(l){5-5} 
% UART & 115.2 kbits/s & 869.752 ms &  & \multirow{3}{*}{1.696 ms} \\ \cmidrule(r){1-3}
% I\textsuperscript{2}C & 400 kbits/s & 226.696 ms &  &  \\ \cmidrule(r){1-3}
% SPI & 8 Mbits/s & 11.696 ms &  &  \\ \cmidrule(r){1-3} \cmidrule(l){5-5} 
% \end{tabular}%
% }
% \begin{flushleft}
% \footnotesize ${}^*$ Transfer times assume per-byte framing: UART 8N1 $\Rightarrow$ 10 bits/byte; I\textsuperscript{2}C (byte + Acknowledgement) $\Rightarrow$ 9 bits/byte; SPI $\Rightarrow$ 8 bits/byte.
% \end{flushleft}
% \end{table}

{\renewcommand{\arraystretch}{1.3}%
\begin{table}[b]
\centering
% \color{blue}
\caption{Detection performance of AdvScan on TinyML benchmarks and target microcontroller platforms against adversarial examples generated by FGSM and PGD-$L_{\infty}$ and C\&W under $L_{0}$, $L_{2}$, and $L_{\infty}$, for both benign and adversarial (AE) inputs. All results reflect AdvScan applied to the model’s final FC layer.}
\label{table:acc}
\resizebox{\columnwidth}{!}{%
\begin{tabular}{@{}ccccc@{}}
\toprule
\textbf{\begin{tabular}[c]{@{}c@{}}Dataset\\ (Model Arc.)\end{tabular}} & \textbf{\begin{tabular}[c]{@{}c@{}}Target\\ Device\end{tabular}} & \textbf{\begin{tabular}[c]{@{}c@{}}MNIST\\ (2 Conv + 1 FC)\end{tabular}} & \textbf{\begin{tabular}[c]{@{}c@{}}CIFAR-10\\ (ResNet-20)\end{tabular}} & \textbf{\begin{tabular}[c]{@{}c@{}}Speech\\ Command\\ (DS-CNN)\end{tabular}} \\ \midrule
\multirow{2}{*}{\begin{tabular}[c]{@{}c@{}}Detection\\ Performance\end{tabular}} & \begin{tabular}[c]{@{}c@{}}STM32F303RC\\ (Arm Cortex-M4)\end{tabular} & \begin{tabular}[c]{@{}c@{}}99.99\%\\ (45,996/46,000)\end{tabular} & \begin{tabular}[c]{@{}c@{}}99.985\%\\ (45,993/46,000)\end{tabular} & \begin{tabular}[c]{@{}c@{}}99.977\%\\ (67,185/67,200)\end{tabular} \\ \cmidrule(l){2-5} 
 & \begin{tabular}[c]{@{}c@{}}STM32L562RE\\ (Arm Cortex-M33)\end{tabular} & \begin{tabular}[c]{@{}c@{}}99.995\%\\ (45,998/46,000)\end{tabular} & \begin{tabular}[c]{@{}c@{}}99.993\%\\ (45,997/46,000)\end{tabular} & \begin{tabular}[c]{@{}c@{}}99.986\%\\ (67,191/67,200)\end{tabular} \\ \bottomrule
\end{tabular}%
}
\end{table}
}

% Efficiency measures AdvScan’s ability to achieve reliable AE detection with minimal runtime overhead. For each inference, AdvScan requires only a single power trace measurement, enabling detection in a reasonable scenario where AEs are only applied to the model a single time. In our evaluation involving 46,000 test inputs (comprising 45,000 adversarial cases generated via FGSM, PGD–$(L_{\infty})$, C\&W with $L_0, L_2, L_{\infty}$ norms, and 1,000 benign images), AdvScan attained near-perfect accuracy (99.99\%) using a user-defined threshold at $P_{th} = 10^{-5}$. This high accuracy required just one trace per inference, which can be collected and processed while the inference is being performed on the edge device. The necessary processing was limited as well, requiring only Pearson correlation and filtering at test time. Hence, AdvScan can reasonably be deployed in a resource constrained scenario and even function at runtime, providing rapid AE detection. These results highlight the lightweight nature of AdvScan, making it suitable for deployment in resource-constrained edge environments.
{
\subsubsection{\bf{Success Criterion 3 (Reliability)}}

Reliability quantifies AdvScan’s detection consistency by examining its false positive (FP) and false negative (FN) rates. A reliable AE detection method ideally exhibits minimal FPs (benign inputs incorrectly flagged as AEs) and FNs (AEs incorrectly flagged as benign). Based on our evaluation in Table \ref{table:acc} and Fig. \ref{fig:fig_7} for $P_{th} = 10^{-5}$, AdvScan exhibits zero observed FPs across all three TinyML benchmarks (MNIST, CIFAR-10, and Speech Commands), both target MCUs, and all considered AE-generation methods (FGSM, PGD-$L_{\infty}$, and C\&W under $L_0$, $L_2$, and $L_{\infty}$ norms). Across the STM32F303RC, we observe a total of 26 FNs (4, 7, and 15 misclassified cases for MNIST, CIFAR-10, and Speech Commands, respectively), while the STM32L562RE exhibits 14 FNs (2, 3, and 9 misclassified cases for the same datasets). In aggregate, this yields only 40 misdetections out of 318,400 total test cases across both boards, i.e., an overall misclassification rate of approximately $0.0126\%$ when counting both benign and AE inputs. These results are obtained with only a single FC-layer power trace per test input on both MCUs (see Sec. \ref{sec:layer}), further minimizing operational delay, as shown in Table \ref{tab:latency}.
% Reliability quantifies AdvScan's detection consistency by examining its false positive (FP) and false negative (FN) rates. A reliable AE detection method ideally exhibits minimal FPs (benign inputs incorrectly flagged as AEs) and FNs (AEs incorrectly flagged as benign). Based on our evaluation, as shown in Table \ref{table:acc} and Fig. \ref{fig:fig_7} for $P_{th}=10^{-5}$, AdvScan resulted in zero observed false positives and only XXX false negatives, corresponding to an FN rate of $XXX\times10^{-X}$\% over adversarial inputs. Additionally, AdvScan can reliably operate with only a single test trace from the fully connected layer per test input, further minimizing operational delay, as shown in Table \ref{tab:latency}.
}
% Reliability quantifies AdvScan's detection consistency by examining its false positive (FP) and false negative (FN) rates. A reliable AE detection method ideally produces minimal FP (benign inputs incorrectly flagged as AEs) and FN (AEs incorrectly flagged as benign) outcomes. Based on our evaluation, as shown in Table. \ref{table:acc} and Fig. \ref{fig:fig_7} with $P_{th}=10^{-5}$, we found that AdvScan resulted in zero observed false positives and only four false negatives which are shown as in red boxes in the table. This corresponds to an FN rate of $8.889\times10^{-5}$\% for adversarial inputs. Additionally, AdvScan operates at the fully connected layer without any model retraining or input preprocessing, so it preserves the original inference accuracy, unlike other defenses that degrade performance through input transformations. 

{
\subsubsection{\bf{Success Criterion 4 (Generalizability)}}

Generalizability assesses AdvScan's performance across diverse TinyML models, AE-generation methodologies, and data modalities and hardware platforms. Figure \ref{fig:fig_7} demonstrates AdvScan’s performance for an STM32F303RC target MCU, as a function of attack strength across multiple TinyML models (DS-CNN, ResNet-20, and a simple CNN architecture) trained on various datasets (speech commands, CIFAR-10, and MNIST), and diverse AE-generation approaches (FGSM, PGD, and C\&W with different norms). Across these models, datasets, and attack configurations, AEs consistently yield $p$-values below the detection threshold ($P_{\text{th}} = 10^{-5}$), while benign inputs remain above this threshold, indicating that AdvScan reliably separates benign and AE behavior across model architectures, attack methodologies, and data modalities on this device. 

Table \ref{table:acc} demonstrates that AdvScan generization can be extended across different hardware platforms. To assess cross-device robustness, we re-ran the evaluation from Fig. \ref{fig:fig_7} on an additional target MCU, the STM32L562RE (Arm Cortex-M33), using the same TinyML models, datasets, and AE-generation methods (FGSM, PGD-$L_{\infty}$, and C\&W under $L_0$, $L_2$, and $L_{\infty}$ norms). For each device, AdvScan evaluated 159{,}200 test inputs (46,000 MNIST, 46,000 CIFAR-10, and 67,200 Speech Commands), comprising both benign inputs and AEs. With the same detection threshold as past experiments, set at $P_{\text{th}} = 10^{-5}$, AdvScan correctly classifies 159,174 inputs on STM32F303RC and 159,186 inputs on STM32L562RE, corresponding to overall detection accuracies of approximately 99.984\% and 99.991\%, respectively. In both cases, per-dataset accuracies remain above 99.97\% (see Table \ref{table:acc}), with only 26 and 14 misdetections on STM32F303RC and STM32L562RE, respectively. Here, AdvScan operates without requiring any knowledge of the underlying AE-generation methodology, relying solely on measurable deviations in power-consumption-based similarity. Collectively, the results indicate AdvScan’s strong generalization capability across model architectures, datasets, AE types, and MCUs.
}
% Generalizability assesses AdvScan's performance across diverse TinyML models, attack methods, and data modalities. Figure \ref{fig:fig_7} demonstrates AdvScan’s performance across multiple TinyML models (DS-CNN, ResNet-20, and a simple CNN architecture) trained various datasets (speech commands, CIFAR-10, and MNIST), and diverse AE-generation approaches (FGSM, PGD, and C\&W with different norms). The consistently high detection accuracy for diverse detection scenarios in Sec \ref{sec:e1}, \ref{sec:layer}, and \ref{sec:all_in} underscores AdvScan’s strong generalization capabilities.

\section{Conclusion}
In this work, we present AdvScan, a runtime detection method for adversarial examples (AEs) targeting TinyML models, which leverages power side-channel analysis to monitor input integrity. {AdvScan operates entirely in a black-box scenario, requiring only a single power trace from per test input and with no access to model internals, making it well suited for resource-constrained and licensing-constrained edge deployments.} The algorithm identifies AEs by employing hypothesis testing (i.e., a one-sample t-test) to detect statistically significant deviations in the instantaneous power consumption caused by adversarial perturbations. {We implemented AdvScan on two target MCUs: an STM32F303RC (Arm Cortex-M4) and an STM32L562RE (Arm Cortex-M33), and evaluated it on three MLPerf Tiny models (DS-CNN, ResNet-20, and a 2-Conv-1-FC network). Across 318,400 test cases and adversarial inputs crafted using FGSM, PGD, and C\&W attacks, AdvScan achieved a detection rate of 99.987\%, with only 40 false negatives and no false positives.} These results highlight AdvScan's potential as a reliable, efficient, and generalizable method for protecting resource-constrained TinyML deployments against AEs.

% In this work, we presented AdvScan, a runtime detection method for adversarial examples (AEs) targeting TinyML models, leveraging power side-channel analysis to monitor input integrity. Unlike existing solutions, AdvScan operates entirely in a black-box scenario and results in no accuracy degradation, making it ideal for accuracy-critical and licensing scenarios. AdvScan identifies AEs by employing hypothesis testing (i.e., a one-sample t-test) to detect statistically significant deviations in the instantaneous power consumption caused by adversarial perturbations. We implemented and evaluated AdvScan on an STM32F303RC MCU (ARM Cortex-M4) across three MLPerf-Tiny models (DS-CNN, ResNet-20, and a 2-Conv-1-FC network), achieving a high AE detection rate for AEs generated by FGSM, PGD, and C\&W attacks. Across 46,153 test cases, AdvScan achieved a detection rate of 99.99\%, producing only four false negatives and zero false positives. These results highlight AdvScan's potential as a reliable, efficient, and generalizable method for protecting resource-constrained TinyML deployments against AEs.

\bibliographystyle{IEEEtran}
\bibliography{New_Reference}

% \begin{IEEEbiography}
% [{\includegraphics[width=1in,height=1.25in]{Authors/rp.jpg}}]{Robi Paul} (Student Member, IEEE) received his B.S. degree in Electrical Engineering from the Shahjalal University of Science and Technology, Bangladesh, in 2023. He is currently pursuing his Ph.D. degree in Computer Engineering at the Rochester Institute of Technology, Rochester, USA. His current research interests include hardware security, with a focus on on anomaly detection using side-channel leakage. 
% \end{IEEEbiography}

% \begin{IEEEbiography}
% [{\includegraphics[width=1in,height=1.25in]{Authors/mz.jpg}}]{Michael Zuzak} (Member, IEEE) received the Ph.D. degree in electrical engineering from the University of Maryland, College Park, MD, USA, in 2022. He is an Assistant Professor with the Department of Computer Engineering, Rochester Institute of Technology, Rochester, NY, USA. His current research interests include hardware security, computer architecture, and electronic design automation.
% \end{IEEEbiography}
\end{document}